\documentclass[aps, prd, twocolumn, nofootinbib, floatfix, superscriptaddress]{revtex4-1}
\usepackage{subfigure}
\usepackage{graphicx}
\usepackage{dcolumn}
\usepackage{epsfig}
\usepackage{amsmath}
\usepackage{amsfonts}
\usepackage{graphicx}
\usepackage{amssymb}
\usepackage{amssymb,amsmath,amsfonts}
\usepackage{xfrac}
\usepackage{color}
\usepackage{tikz}
\usepackage[utf8]{inputenc}

\usepackage{tcolorbox}
\usepackage{hyperref}
\hypersetup{colorlinks, citecolor=bluscuro, linkcolor=black, urlcolor=bluscuro}
\definecolor{rossos}{cmyk}{0,1,1,0.55}
\definecolor{bluscuro}{rgb}{0.15, 0.2, .85}
\definecolor{bluchiaro}{cmyk}{1,.3,0.,0.1}

\let\oldsqrt\sqrt
\def\sqrt{\mathpalette\DHLhksqrt}
\def\DHLhksqrt#1#2{%
\setbox0=\hbox{$#1\oldsqrt{#2\,}$}\dimen0=\ht0
\advance\dimen0-0.2\ht0
\setbox2=\hbox{\vrule height\ht0 depth -\dimen0}%
{\box0\lower0.4pt\box2}}

\newcommand{\sss}[1]{{\scriptscriptstyle{#1}}}

\newcommand{\uPl}{\mathrm{Pl}}

\newcommand{\usssPl}{\sss{\uPl}}

\newcommand{\Mp}{M_\usssPl}

\newcommand{\beq}{\begin{equation}}
\newcommand{\eeq}{\end{equation}}
\newcommand{\bea}{\begin{equation}\begin{aligned}}
\newcommand{\eea}{\end{aligned}\end{equation}}

\newlength{\wsingfig}
\newlength{\wdblefig}
\newlength{\wquadfig}
\newlength{\wtriplefig}
\newcommand{\Eq}[1]{Eq.~(\ref{#1})}

\newcommand{\Fig}[1]{Fig.~{\ref{#1}}}

\newcommand{\Sec}[1]{Sec.~\ref{#1}}

\newcommand{\App}[1]{Appendix~\ref{#1}}

\newcommand{\be}{\begin{equation}}

\begin{document}

\title{Constraining $F(R)$ bouncing cosmologies through primordial black holes }

\author{Shreya Banerjee}
\email{shreya.banerjee@fau.de}
\affiliation{\mbox{Institute for Quantum Gravity, FAU Erlangen-Nuremberg, Staudtstr. 7, 91058 Erlangen, Germany}}

\author{Theodoros Papanikolaou}
\email{theodoros.papanikolaou@noa.gr}
\affiliation{ \mbox{National Observatory of Athens, Lofos Nymfon, 11852 Athens, Greece} }

\author{Emmanuel N. Saridakis}
\email{msaridak@noa.gr}
\affiliation{ \mbox{National Observatory of Athens, Lofos Nymfon, 11852 Athens, Greece} }
\affiliation{CAS Key Laboratory for Researches in Galaxies and Cosmology, 
Department of Astronomy, \\
University of Science and Technology of China, Hefei, 
Anhui 230026, P.R. China}
\affiliation{\mbox{Departamento de Matem\'{a}ticas, Universidad Cat\'{o}lica del 
Norte, 
Avda.
Angamos 0610, Casilla 1280 Antofagasta, Chile}}

\begin{abstract} 
The phenomenology of primordial black hole (PBH) physics and the
associated PBH abundance constraints, can be used in order to probe the physics of the
early Universe. In this work, we investigate the PBH formation during the standard
radiation-dominated era by studying the effect of an early F(R) modified gravity phase
with a bouncing behavior which is introduced to avoid the initial spacetime singularity
problem. In particular, we calculate the energy density power spectrum at horizon
crossing time and then we extract the PBH abundance in the context of peak theory as
a function of the parameter $\alpha$ of our $F(R)$ gravity bouncing model at hand. Interestingly,
we find that in order to avoid GW overproduction from an early PBH dominated era before Big Bang Nucleosynthesis (BBN), $\alpha$ should lie within the range $\alpha\leq 10^{-19}\Mp^2$. This constraint can be translated to a constraint on the energy scale at the onset of the Hot Big Bang (HBB) phase, $H_\mathrm{RD}\sim \sqrt{\alpha}/2$ which can be recast as $H_\mathrm{RD}< 10^{-10}\Mp$.
\noindent

\end{abstract}

\maketitle

\section{Introduction}

The theory of 
inflation~\cite{Starobinsky:1980te,Guth:1980zm,Linde:1981mu,Albrecht:1982wi,
Linde:1983gd} constitutes a very promising paradigm to account for the physical conditions that prevailed in early universe, being able to address a number of cosmological issues like the horizon and the flatness problems.  However, inflationary theories face the problem of initial 
singularity~\cite{Borde:1996pt}. One attractive alternative  to inflation is the non-singular bouncing  
cosmological paradigm~\cite{PhysRevLett.68.1969,Brandenberger:1993ef}, which assumes that the universe existed forever before the HBB era in a contracting phase and at some point transitioned into the expanding universe that we observe today. Apart from solving the singularity problem the  bounce
realization  can also address the usual flatness and horizon 
problems of standard Big Bang cosmology (for a review on bouncing cosmologies, 
see \cite{Novello:2008ra}) and give rise to an observationally compatible 
cosmological power spectrum \cite{LILLEY20151038, Battefeld:2014uga, 
PhysRevD.78.063506}. 

In order to acquire a non-singular bouncing phase,  violation of the null 
energy condition is necessary. Consequently, modified 
gravity theories \cite{CANTATA:2021ktz,Nojiri:2006ri,Capozziello:2011et,Benisty:2021sul,Benisty:2021laq}  
provide an ideal framework for obtaining a bouncing universe. Hence, such 
bouncing solutions have been constructed through various approaches to modified 
gravity, such as the Pre-Big-Bang~\cite{Veneziano:1991ek} 
and the Ekpyrotic~\cite{Khoury:2001wf,Khoury:2001bz} models, gravitational 
theories  whose gravity actions contain higher 
order corrections~\cite{Biswas:2005qr,Nojiri:2013ru}, $F(R)$ 
gravity~\cite{Bamba:2013fha,Nojiri:2014zqa}, $f(T)$ 
gravity~\cite{Cai:2011tc} models, braneworld scenarios~\cite{Shtanov:2002mb,Saridakis:2007cf}, 
non-relativistic gravity~\cite{Cai:2009in,Saridakis:2009bv},  massive 
gravity~\cite{Cai:2012ag}, etc. The above scenarios can be further extended to 
the paradigm of cyclic 
cosmology~\cite{Lehners:2008vx,Banerjee:2016hom,Saridakis:2018fth}.

As a potential candidate,  the bounce
scenario is expected to be consistent with current cosmological observations and 
to be distinguishable from the experimental predictions of cosmic inflation as 
well as other paradigms~\cite{Cai:2014bea,Cai:2014xxa}. One interesting way to 
constrain such bouncing scenarios is the study of their effect on the formation of 
primordial black holes (PBHs)~\cite{Carr:1975qj, Carr:2009jm}.

Primordial black holes, first proposed in early '70s ~\cite{1967SvA....10..602Z, Carr:1974nx,1975ApJ...201....1C}, are considered to 
form in the very early universe out of the gravitational collapse of very high 
overdensity regions, whose energy density is higher than a critical 
threshold~\cite{Harada:2013epa,Musco:2018rwt,Kehagias:2019eil,Musco:2020jjb,
Musco:2021sva,Addazi:2021xuf,Papanikolaou:2022cvo}. According to recent arguments, PBHs can 
naturally act as a viable dark matter candidate 
~\cite{Chapline:1975ojl,Clesse:2017bsw} and potentially explain the generation 
of large-scale structures through Poisson 
fluctuations~\cite{Meszaros:1975ef,Afshordi:2003zb}, while they can also seed 
the supermassive black holes residing in galactic centres 
~\cite{1984MNRAS.206..315C, Bean:2002kx}. Furthermore,   they 
are associated with numerous gravitational-wave (GW) signals, from black-hole 
merging events~\cite{Nakamura:1997sm, Ioka:1998nz, 
Eroshenko:2016hmn,Zagorac:2019ekv, Raidal:2017mfl} up to primordial 
second-order scalar induced GWs from primordial curvature 
perturbations~\cite{Bugaev:2009zh, Saito_2009, Nakama_2015, 
Yuan:2019udt,Zhou:2020kkf,Fumagalli:2020nvq} (for a recent 
review see \cite{Domenech:2021ztg}) or from Poisson PBH energy density 
fluctuations~\cite{Papanikolaou:2020qtd,Domenech:2020ssp,Kozaczuk:2021wcl}.  
Other indications in favor of the PBH scenario can be found in 
~\cite{2018PDU....22..137C}.
Their abundance is constrained from a wide variety of 
probes~\cite{Carr:2017jsz,Kuhnel:2017pwq,Bellomo:2017zsr,Clesse:2017bsw,
Green:2014faa,Sasaki:2018dmp} over a range of masses from $10\mathrm{g}$ up to 
$10^{20}M_\odot$, thus giving us access to a very rich phenomenology.

Up to now, the majority of the literature studied PBH formation within 
single-field~\cite{Garcia-Bellido:2017mdw,Motohashi:2017kbs,Ezquiaga:2017fvi,
Martin:2019nuw} or 
multi-field~\cite{Clesse:2015wea,Palma:2020ejf,Fumagalli:2020adf} inflationary 
cosmology. It was also studied within modified theory set-ups~\cite{Kawai:2021edk,Yi:2022anu,Zhang:2021rqs}. However, the study of PBHs in bouncing scenarios is 
limited~\cite{Carr:2011hv, 
Carr:2014eya,Quintin:2016qro,Chen:2016kjx,Clifton:2017hvg}, most of which has 
been done with a generalised approach, without any falsification of the bouncing 
scenarios. Therefore, given the aforementioned rich phenomenology and the 
associated PBH abundance constraints over a range of masses which span more than 
$50$ orders of magnitude, PBHs can clearly provide a novel promising way to test 
and constrain various bounce scenarios.

In this work, we investigate the bounce realization within one of the simplest 
modifications of general relativity which can violate the null energy condition 
and thus  give rise to a bouncing phase, namely the $F(R)$ gravity theory. 
$F(R)$ gravity   forms a particular class of theories in which the 
Einstein–Hilbert action   is upgraded to a general function of the Ricci scalar 
$R$ ~\cite{Nojiri:2006ri}. $F(R)$ theories have been studied extensively in 
the context of inflation \cite{Inagaki:2019hmm,Nojiri:2007as,Nojiri:2017qvx}, 
bounce \cite{Odintsov:2020zct,Bamba:2013fha,Nojiri:2014zqa} and late-time 
acceleration \cite{Hu:2007nk,Odintsov:2020zct,Carroll:2003wy}. Additionally,  
this class of theories has been highly successful in explaining both late and 
early time acceleration along with the intermediate thermal history of the 
Universe (see \cite{DeFelice:2010aj, Nojiri:2010wj} for   reviews). Therefore, 
it would be very interesting to examine how such theories can be constrained or 
ruled out through the study of PBH formation within them.

The manuscript is organised as follows: In \Sec{sec:f(R)} we introduce a class of $F(R)$ gravity theories which can induce a bouncing  scale factor. Then, in \Sec{sec:Perturbations} we extract the curvature power spectrum close to the bounce as a function of the theoretical parameters evolved, namely the bouncing parameter $\alpha$, matching it to the curvature power spectrum during the standard radiation era when PBHs are assumed to form. Subsequently, in \Sec{sec:PBH}, we present the formalism to compute the PBH mass function $\beta(M)$ within peak theory. Followingly, in \Sec{sec:results} after investigating the effect of an initial $F(R)$ gravity phase close to the bounce on the curvature power spectrum $\mathcal{P}_\mathrm{\delta}(k)$ and the PBH mass function $\beta(M)$ we set constraints on 
$\alpha$ by requiring that GWs induced from PBH Poisson fluctuations during an early PBH dominated era before BBN are not overproduced. Finally, \Sec{sec:conclusions} is devoted to conclusions.


\section{Bounce cosmology through $F(R)$ gravity}\label{sec:f(R)}

For the present analysis we   consider the  flat 
Friedman-L\^emaitre-Robertson-Walker (FLRW) background metric 
\begin{equation}
 \mathrm{d}s^2 = - \mathrm{d}t^2 + a^2(t)\delta_{ij}\mathrm{d}x^i\mathrm{d}x^j,
\end{equation}
where $a(t)$ is the scale factor while the gravitational action for $F(R)$ gravity in vacuum can be written as:
\begin{eqnarray}
\begin{split}
 S & = \frac{1}{2\kappa^2} \int  \mathrm{d}^4x \sqrt{-g} F(R)  \\ & = \frac{1}{2\kappa^2} \int  \mathrm{d}^4x \sqrt{-g} R+\frac{1}{2\kappa^2} \int  \mathrm{d}^4x \sqrt{-g} f(R),
 \label{basic1}
 \end{split}
\end{eqnarray}
where $\kappa^2 = 8\pi G = \frac{1}{M_\mathrm{Pl}^2}$, with $M_\mathrm{Pl}$ being the reduced Planck mass. Here, we choose $F(R)=R+f(R)$, with the function $f(R)$ capturing deviation effects from General Relativity (GR). In the following, we assume that the terms coming from the function $f(R)$ have considerable contributions in and around the bounce. This is because we introduce this extra $f(R)$ function at the level of the gravitational action in order to account for the problem of the initial spacetime singularity. On the other hand, as we move away from the bounce into the standard radiation-dominated (RD) era, we gradually switch-off the $f(R)$ contribution and the action reduces to that of GR, given also its very good agreement with the current cosmological data up to the era of Big Bang Nucleosynthesis. 

We proceed now to the reconstruction of the $f(R)$ function close to the bounce. The corresponding Friedmann 
equations close to the bounce turn out to be
\beq
\begin{split}
3H^2  = -\frac{f(R)}{2} + & 3\big(H^2 + \dot{H}\big)f'(R) \\ & -  18\big(4H^2\dot{H} + 
H\ddot{H}\big)f''(R) \label{basic4}
\end{split}
\eeq
\beq\label{basic4bb}
\begin{split}
\frac{f(R)}{2}& = \big(3H^2 + \dot{H}\big)f'(R) \\ & - 6\big(8H^2\dot{H} + 4\dot{H}^2 + 6H\ddot{H}
+ \dddot{H}\big)f''(R) \\
& - 36\big(4H\dot{H} + \ddot{H}\big)^2f'''(R),
\end{split}
\eeq
where $H(t)\equiv \dot{a}/a$ is the Hubble parameter.

Since we are interested in studying the bounce realization within  $F(R)$ gravity, we choose the scale factor accordingly. The general evolution of the universe in bouncing cosmology consists of a period of contraction followed by a cosmological bounce and then by the standard expanding universe. Any form of the scale factor satisfying $a(t_b)> 0,\ \dot{a}(t_b)=0,\ \ddot{a}(t_b)> 0$, is capable for giving rise to a bouncing cosmology, where $t_b$ corresponds to the time when the bounce occurs.

Let us now present the bounce realization at the background level. Without loss of generality we consider a bouncing scale factor of the form
\begin{eqnarray}\label{eq:a_close_to_the_bounce}
 a_b(t) = 1 + \alpha t^2,
 \label{rec_bounce scale}
\end{eqnarray}
 with
$\alpha$ being a free parameter and the
bounce happening at $t = 0$. The above form of scale factor has been obtained by 
keeping terms up to quadratic order in $t$ in the Taylor   expansion of $a(t)$ 
near the bounce. We neglect higher order terms as we are interested for 
solutions near the bounce. Finally, note that the bounce realization conditions  mentioned above indicate  that $\alpha>0$. For different parametrisations of the scale factor close to the bounce see \App{app:different_a_parametrisations}.

Using the above form of the scale factor, we obtain the expressions for the 
Hubble parameter and the Ricci scalar (keeping   terms up to 
$\mathcal{O}(\alpha t^2)$) as:
\beq
\begin{split} 
H(t)&=\frac{2\alpha t}{1 + \alpha t^2} \simeq 2\alpha t, \\
R(t)&=12H^2 + 6\dot{H} = \frac{12\alpha(1 + 3\alpha t^2)}{(1 + \alpha t^2)^2} 
\\ & \simeq 12\alpha + 12\alpha^2t^2.
\end{split}
\eeq
As we can see from the above relations, the Hubble parameter varies linearly 
with time around the bounce, and becomes zero at the bounce point, as expected. 
Moreover, the Ricci scalar at the bounce is $R(0)=12\alpha$. Inserting the  
above expressions into \Eq{basic4} we acquire
\beq
\begin{split}
24\alpha(R - 12\alpha)f_b''(R) + & (R - 24\alpha)f_b'(R) \\ & + f_b(R) + 2(R - 12\alpha) = 0,\label{rec_bounce eom}
\end{split}
\eeq
 where the index $b$ refers to background quantities. Finally, solving the above equation for $f_b(R)$ and keeping terms up to $\mathcal{O}(\alpha t^2)$, the solution for $F_b(R)$ near the bounce can be recast as~\cite{Odintsov:2020zct}
 \beq\label{eq:F_R_bounce}
 \begin{split}
F_b=R+ & e^{-\frac{R}{24\alpha}}\Big(\frac{12\alpha-C}{216\alpha}\Big)\Biggl[12e^{\frac{R}{24\alpha}}R \\ & +\sqrt{\frac{6e\pi}{\alpha}}(R-12\alpha)^{3/2}\mathrm{Erfi}\left(\sqrt{\frac{R-12\alpha}{24\alpha}}\right)\Biggr],\\
  \end{split}
 \eeq
 where $\mathrm{Erfi}(z)$ is the imaginary error function defined as 
$\mathrm{Erfi}(z) = -i\mathrm{Erf}(iz)$ and $C$ is an integration constant which 
will be fixed later. Hence, from now on the parameter $\alpha$ can be 
considered as the $F(R)$ model parameter. 

The form of $F(R)$ obtained above is valid in and around the bounce i.e. in the region where the form of the scale factor is given by Eq. \eqref{rec_bounce scale} with $\alpha t^2\lesssim 1$. For this reason, in the following we will naturally consider that the transition to the RD era, where one recovers the standard GR evolution, happens around the time when the perturbative expansion of the scale factor in \Eq{eq:a_close_to_the_bounce} breaks down, namely when $\alpha t^2 \sim 1$. Consequently, one gets that $t_\mathrm{RD}$ is given by 
\beq\label{eq:t_RD}
t_\mathrm{RD} \sim \frac{1}{\sqrt{\alpha}}.
\eeq

\begin{figure}\centering
{\includegraphics[width=0.8\linewidth]{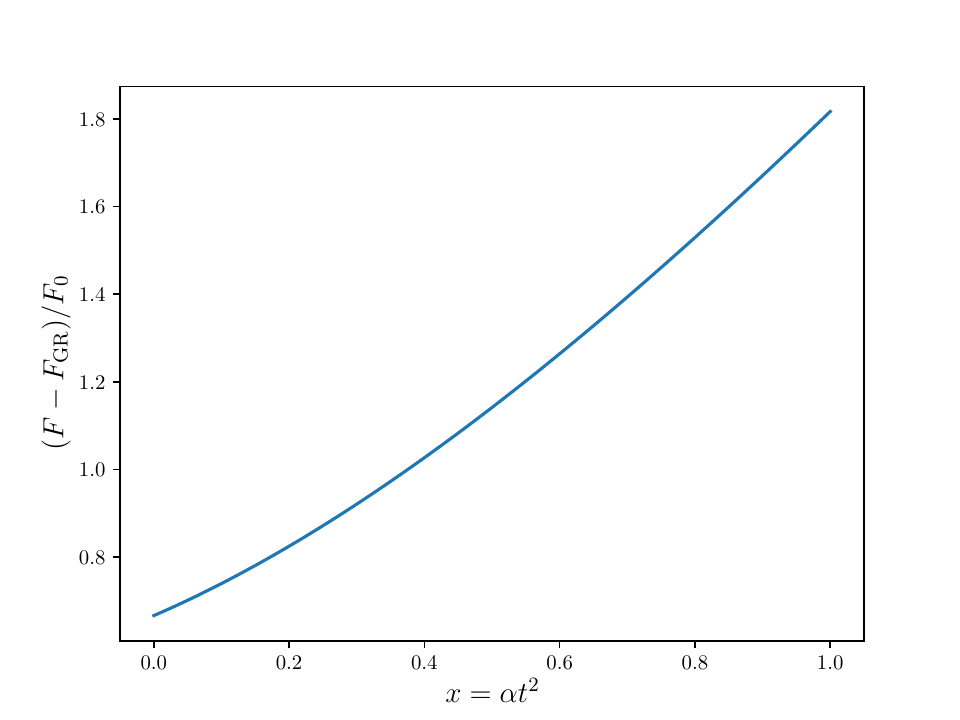}\\
\includegraphics[width=0.8\linewidth]{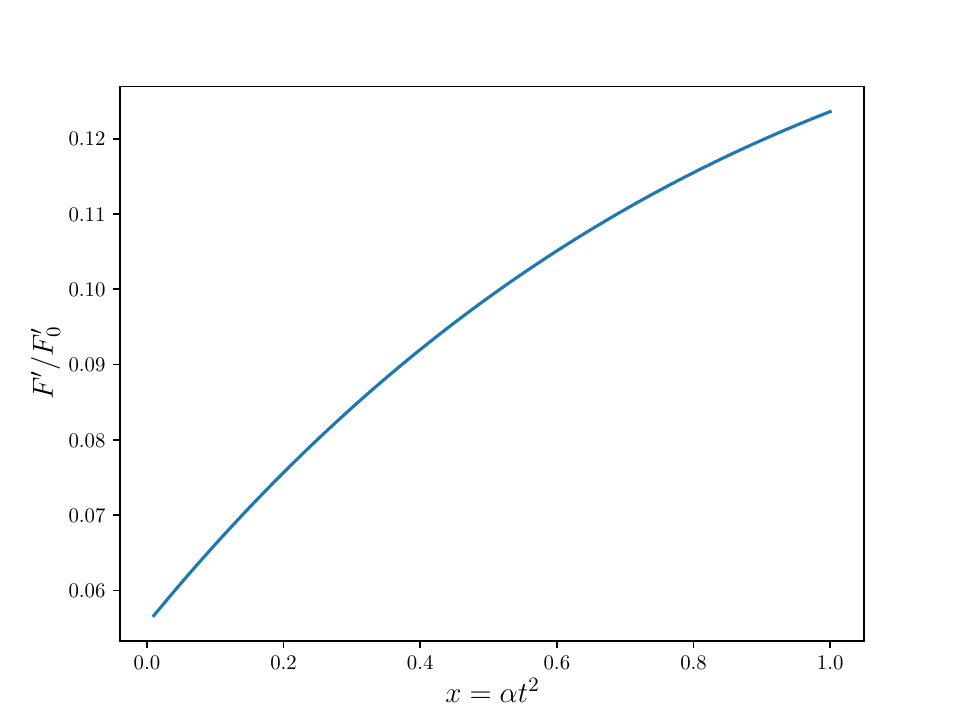}\\
\includegraphics[width=0.8\linewidth]{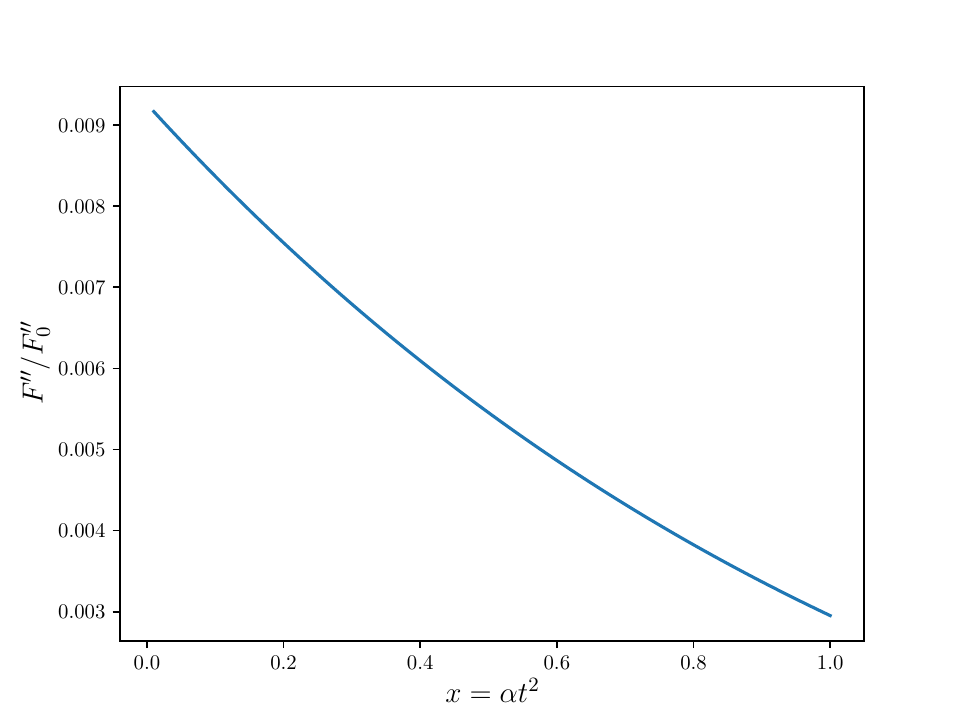}
}
\caption{\it{The functions $F$ (upper graph), $F^\prime$ (middle graph) and $F^{\prime\prime}$ (lower graph), in terms of the time variable $x$ defined as $x\equiv \alpha t^2$, with $F_\mathrm{GR}=R$ and $F_0 = (12\alpha - C)$,$F^\prime_0 = (12\alpha - C)/alpha$ and $F^{\prime\prime}_0 = (12\alpha - C)/\alpha^2$. }}
\label{fig:viable_F_R_model_F}
\end{figure}

Before deriving in the next section the comoving curvature perturbation within our $F(R)$ bouncing model we need to make here an instability analysis of the underlying gravity theory close to the bounce. In particular, in order to avoid ghosts~\cite{DeFelice:2006pg},  the first derivative of the function $F(R)$ should be positive, i.e. $F^\prime \equiv \partial F/\partial R > 0$ while at the same time, in order to avoid tachyonic instabilities, the square of the mass of scalaron field $M^2$, where $M^2 \sim 1/F^{\prime\prime}$ with $F^{\prime\prime} \equiv \partial^2 F/\partial R^2$, should be positive~\cite{DeFelice:2010aj}.
These in turn arise from the perturbation analysis of the theory performed in \cite{Amendola:2006we,Starobinsky:2007hu}, and in particular from the comoving curvature perturbation $\mathcal{R}$, under the requirement to have a successful cosmological evolution from radiation era till matter domination.
Thus, the conditions for a viable $F(R)$ bouncing model are the following:
\beq\label{eq:viable_F_R_model}
F^\prime>0\quad\mathrm{and}\quad F^{\prime\prime}>0.
\eeq
From \Eq{eq:F_R_bounce} one can derive $F^\prime$ and $F^{\prime\prime}$ which can be recast as 
\beq\label{eq:F_R_F_RR}
\begin{split}
F'\left[R(t)\right]&=\frac{ (12\alpha-C)}{36t\alpha^2} \Bigl[2t\alpha+t^3 \alpha^2 \\ & +\sqrt{2\alpha} t^2\alpha (3-t^2\alpha)F_\mathrm{D}\left(\frac{t\sqrt{\alpha}}{\sqrt{2}}\right)\Bigr],
\end{split}
\eeq
\beq
\begin{split}
F''\left[R(t)\right] & =  \frac{(12\alpha-C)}{864\alpha^3 t^2}\Bigl\{\alpha t^2(5-\alpha t^2) + \sqrt{2\alpha t^2}[3 \\ & + \alpha t^2 (\alpha t^2 -6) ]F_\mathrm{D}\left(\frac{t\sqrt{\alpha}}{\sqrt{2}}\right)\Bigr\},
\end{split}
\eeq
 where $F_\mathrm{D}(x)$ is the Dawson function.
Below, we plot the functions $F$, $F^\prime$ and $F^{\prime\prime}$ as a function of time, by using $x\equiv \alpha t^2$ as the time variable. Thus, we reach times up to $x=1$ when the perturbative expansion of the scale factor in \Eq{eq:a_close_to_the_bounce} breaks down and one enters the standard RD era as explained before. We choose the value of the integration constant $C$ to be such as that $C<12\alpha$ so that the conditions in \ref{eq:viable_F_R_model} are satisfied. As it can be seen from Fig. \ref{fig:viable_F_R_model_F},  for $C<12\alpha$ the conditions \Eq{eq:viable_F_R_model} are satisfied making our $F(R)$ bouncing model free of ghosts and tachyonic instabilities.

\section{The curvature power spectrum}\label{sec:Perturbations}

Since we have  studied  in the previous section the background behavior of a bouncing scenario realized 
within $F(R)$ gravity and we have extracted the function $F(R)$ around the 
bounce, we proceed to the calculation of the curvature power spectrum by 
deriving the corresponding comoving curvature perturbation.

\subsection{The curvature perturbation}

Before launching our calculation, we  should examine  which  primordial 
perturbation modes are relevant for present-day observation. As we saw 
above, the Hubble parameter vanishes at the bounce point, thus giving rise to 
an infinite comoving Hubble radius ($1/aH$) there. In the following, we 
match the bouncing phase with the standard Hot Big Bang radiation phase, which 
in turn, according to the standard cosmological evolution as dictated by the 
current cosmological probes, is connected to a matter epoch and then at late 
times with an accelerated expansion phase. Consequently, the Hubble horizon 
decreases and tends  to zero for late times, while for cosmic times near 
the bouncing point the Hubble horizon has an infinite size. Therefore, all the 
perturbation modes at that time are contained within the horizon, and at later 
epochs they cross the Hubble radius becoming relevant for current observations. 
Hence, in the following we focus on  the perturbation equations near the 
bounce, namely near $t=0$.

Choosing to work in the comoving gauge, the spatial part of the perturbed scalar metric tensor reads as
 \begin{eqnarray}
 \delta g_{ij} = a^2(t)\big[1 - 2\zeta(\vec{x},t)]\delta_{ij},
 \label{scalar per metric}
\end{eqnarray}
where $\zeta(\vec{x},t)$ denotes the comoving curvature perturbation. The corresponding 
action for the scalar perturbations  
reads as~\cite{Hwang:2005hb,Noh:2001ia,Hwang:2002fp}
\begin{eqnarray}
  \delta S_{\zeta} = \int \mathrm{d}t \mathrm{d}^3\vec{x} a(t) z(t)^2\left[\dot{\zeta}^2
 - \frac{1}{a^2}\left(\partial_i\zeta\right)^2\right]\, ,
 \label{scalar per action}
 \end{eqnarray}
 with $z(t)$ given by the following expression~\cite{Odintsov:2020zct}:
 \begin{eqnarray}
  z(t) = \frac{a(t)}{\kappa\bigg[H(t) + 
\frac{1}{2F'(R)}\frac{dF'(R)}{dt}\bigg]} 
\sqrt{\frac{3}{2F'(R)}\bigg[\frac{dF'(R)}{dt}\bigg]^2}.
  \label{scalar per z}
 \end{eqnarray}
Using the solution for $F(R)$, i.e. \Eq{eq:F_R_bounce}, the expression for $\mathrm{d}F'(R)/\mathrm{d} t$  where $^\prime$ denotes differentiation with respect to the Ricci scalar, is given by
\begin{eqnarray}
  && \frac{\mathrm{d}F'(R(t))}{\mathrm{d}t}= 
\frac{t(12\alpha-C)\left\{t^2\alpha(5-t^2\alpha)\right\}}{
36t^2\alpha} \nonumber\\ && +\frac{t(12\alpha-C)\left\{\sqrt{2\alpha}t [3 +t^2 
\alpha(-6+t^2 
\alpha)]F_\mathrm{D}\left(\frac{t\sqrt{\alpha}}{\sqrt{2}}\right)\right\}}{
36t^2\alpha}. \nonumber \\
\end{eqnarray}
$F'(R)$ is given by \Eq{eq:F_R_F_RR}.

As mentioned earlier, the perturbation modes are generated close to the bounce, 
 therefore  we solve the above equation for cosmic times near the bouncing 
point. As a result, we keep terms upto $\mathcal{O}(\alpha t^2)$ for the rest of our 
analysis. The corresponding expression for $z(t)$, keeping terms up to 
$\mathcal{O}(\alpha t^2)$ in $F'(R)$ and $\frac{\mathrm{d}F'(R(t))}{\mathrm{d}t}$, becomes
 \begin{equation}
     z(t)=\frac{(1/\alpha)^{3/2}\alpha \sqrt{12\alpha-C}}{3^{1/2}(t^2+1)\kappa}+\frac{2\alpha^2\sqrt{12\alpha-C} t^2}{3^{1/2}4\alpha^{3/2}\kappa}.
     \label{z}
 \end{equation}
  At the end, the perturbed action leads to the following Lagrange equation for 
the Fourier mode of the comoving curvature perturbation, $\zeta_k$:
  \begin{eqnarray}
  \frac{1}{a(t)z^2(t)}\frac{d}{dt}\bigg[a(t)z^2(t)\dot{\zeta}_k\bigg] + \frac{k^2}{a^2}\zeta_k(t) = 0.
  \label{scalar per eom2}
 \end{eqnarray}
 In the above equation, by using \eqref{z} and keeping terms upto 
$\mathcal{O}(\alpha t^2)$, the quantity $a(t)z(t)^2$ becomes:
 \begin{eqnarray}
     a(t)z(t)^2=U+V t^2, 
 \end{eqnarray}
 with $U=\frac{(12\alpha-C)}{12\alpha \kappa^2}$, $V=\frac{(12\alpha-C)}{4\kappa^2}$.
 
At the end, the Lagrange equation for $\zeta_k$ can be recast at leading order as 
\begin{eqnarray}
 \ddot{\zeta}_k + \frac{2V}{U}t\dot{\zeta}_k + k^2\zeta_k(t) = 0,
 \label{scalar per eom3}
\end{eqnarray}
whose solution is
\begin{eqnarray}\label{eq:zeta_k}
 && 
 \!\!\!\!\!\!\!\!\!\!\!\!\!\!\!\!\!
 \zeta_k(t) = C_1(k)~e^{-\frac{V}{U}t^2}~H\left(-1 + \frac{k^2U}{2V}, 
\sqrt{\frac{V}{U}}~t\right)
\nonumber \\
&& \  \ +C_2(k)~e^{-\frac{V}{U}t^2}~{}_1F_1\left(\frac12 - 
\frac{k^2U}{4V},\frac12, \sqrt{\frac{V}{U}}~t^2\right),  
 \label{scalar per sol1}
\end{eqnarray}
 where $C_1(k),\ C_2(k)$ are   integration constants, $H(n,x)$ is the n-th 
order Hermite polynomial, and ${}_1F_1(a,b,x)$ is the Kummer confluent 
hypergeometric function.

The expressions for the integration constants $C_1(k), \ C_2(k)$ are obtained by 
setting the initial conditions for the curvature perturbations.  Given the fact 
that close to the bounce the Hubble radius is infinitely large as mentioned 
above, the primordial modes are well inside the Hubble radius thus satisfying 
the condition $k \gg aH$. Therefore, the initial conditions for $\zeta_k$ will 
be set through the Mukhanov-Sasaki variable, defined in the present context as 
$v_k(t) \equiv z(t)\zeta_k(t)$~\cite{Odintsov:2020zct}, and whose value on 
sub-Hubble scales is set by the Bunch-Davies vacuum state, i.e.
\beq\label{eq:Bunch_Davis}
 v_{k,k\ll aH} = \frac{e^{-ik\eta}}{\sqrt{2k}},
\eeq
where the time variable $\eta$ is the conformal time defined by 
$\mathrm{d}\eta \equiv \mathrm{d}t/a(t)$. Using the expression 
\eqref{eq:a_close_to_the_bounce} for the scale factor near the bounce, we 
obtain from \Eq{eq:Bunch_Davis} that 
\beq
\eta = \int^t_{0}\mathrm{d}t^\prime/a(t^\prime) = \frac{\arctan (\sqrt{\alpha} t)}{\sqrt{\alpha}}.
\eeq
 Consequently, the initial conditions satisfied by $v_k$ and its derivative become:
\begin{eqnarray}
 v_k(t\rightarrow 0)=\frac{1}{\sqrt{2k}},\nonumber \\
 \dot{v}_k(t\rightarrow 0)=-\frac{ik\sqrt{\alpha}}{\sqrt{2k}}.
 \label{initial}
\end{eqnarray}

Using these conditions and the fact that $\dot{z}(t\rightarrow 0)=0$,  we finally 
acquire straightforwardly the expressions for the integration constants $C_1,\ 
C_2$ as
\beq
C_1(k)=\frac{3i\kappa 2^{\frac52-\frac{k^2}{6\alpha}}\sqrt{k}\alpha^{3/2}\Gamma\left(\frac32-\frac{k^2}{12\alpha}\right)}{\sqrt{\pi}(6\alpha - k^2)\sqrt{12\alpha - C}}
\eeq
\beq
\begin{split}
C_2(k) &= \frac{\sqrt{2}\kappa}{k^{1/2}(6\alpha - k^2)\sqrt{12\alpha-C}\Gamma\left(1-\frac{k^2}{12\alpha}\right)} \Bigl[ -6i k\alpha^{3/2} \\ & \times\Gamma\left(\frac32-\frac{k^2}{12\alpha}\right)  +\sqrt{3\alpha}\left(6\alpha - k^2\right)\Gamma\left(1-\frac{k^2}{12\alpha}\right)\Bigr],
\end{split}
\eeq
where $\Gamma(x)$ denotes the Gamma function.
At the end, the corresponding curvature power spectrum can be recast as follows:
\begin{eqnarray}
\label{eq:P_zeta_close_to_the_bounce}
\!\!\!\!\!\!\!\!\!\!\!\!\!\!\!\!\!\!\!\!\!\!\!\!\!\!\!\!\!
\begin{split}
\mathcal{P}_{\zeta}(k, t) & \equiv\frac{k^3}{2\pi^2}~\bigg|\zeta_k(t)\bigg|^2 \\ & =\frac{k^3}{
2\pi^2}~\bigg|C_1(k)~e^{-\frac{V}{U}t^2}~H\bigg[-1 + \frac{k^2U}{2V}, 
\sqrt{\frac{V}{U}}~t\bigg]
\\ & 
+C_2(k)~e^{-\frac{V}{U}t^2}~{}_1F_1\bigg[\frac12 - 
\frac{k^2U}{4V},\frac12, \sqrt{\frac{V}{U}}~t^2\bigg]\bigg|^2.  
\end{split}
\end{eqnarray}

\subsection{Matching the bounce with a radiation-dominated era}
As explained in \Sec{sec:f(R)}, close to the bounce the underlying gravity theory is described by a $F(R)$ modified gravity setup with $F(R)$ given by \Eq{eq:F_R_bounce}. During this phase, the scale factor evolution is dictated by \Eq{eq:a_close_to_the_bounce}, which is nothing else than a perturbative expansion close to the bounce, valid for $\alpha t^2\lesssim 1$, and corresponds to a fluid dominated Universe with an equation-of-state parameter
$w=-2/3$. Then, $F(R)$ gravity modifications are switched off and one recovers the standard HBB phase which is described by GR. Consequently, matching the two phases and requiring continuity of the scale factor at the onset of the RD era one gets that 
\begin{equation}\label{eq:matching_bounce_with_RD}
    a(t)=
    \begin{cases}
    1+\alpha t^2 \mathrm{\;,\;}t<t_\mathrm{RD}\\
    a_\mathrm{RD}\left(\frac{t}{t_\mathrm{RD}}\right)^{1/2} \mathrm{\;,\;}t>t_\mathrm{RD},
    \end{cases}
\end{equation}
with $t_\mathrm{RD}$ being the transition time between the exotic 
phase close to the bounce with $w=-2/3$ and the RD phase given by \Eq{eq:t_RD}, and $a_\mathrm{RD}$ the respective 
scale factor at the onset of the RD era. We mention that in order to keep the scale factor continuous 
during the transition we choose $a_\mathrm{RD}$ to be $a_\mathrm{RD}=1+\alpha
t^2_\mathrm{RD}$. 

Given the fact that in the following we elaborate  the power spectrum at 
the horizon crossing time during the RD era, i.e. $k=a(t) H(t)$ with 
$t>t_\mathrm{RD}$, one can find the horizon crossing time 
$t_\mathrm{HC}(k,\alpha)$ by solving $k=a H$ with  
$a(t)=a_\mathrm{RD}\left(\frac{t}{t_\mathrm{RD}}\right)^{1/2}$ and 
$H(t)=\frac{1}{2t}$. At the end, we extract that
\beq\label{eq:t_HC}
t_\mathrm{HC}(k,\alpha)=\frac{\sqrt{\alpha}}{k^2}.
\eeq

At this point it is important to stress out that in the expression
(\ref{eq:P_zeta_close_to_the_bounce}) we derived the curvature power spectrum close to 
the bounce by parametrizing the scale factor as in 
\Eq{eq:a_close_to_the_bounce}. \Eq{eq:a_close_to_the_bounce} describes actually quite well the background dynamical evolution up to the onset of the RD era when the perturbative expansion of the scale factor breaks down. Hence, one can compute 
$\mathcal{P}_{\zeta}(k,t)$ at horizon exiting time during the initial $F(R)$ gravity phase before 
the RD era, namely when $k = a(t) H(t)$ with $t<t_\mathrm{RD}$. 
At this point, we need to stress that in general within the context of bouncing cosmologies, as we pass from the contraction to the expansion phase the comoving curvature perturbation $\zeta_k$ is not necessarily conserved~\cite{Battefeld:2014uga}. However, for non-singular bouncing scenarios as the one we consider here one finds a non-singular evolution of $\zeta_k$ through the bounce~\cite{Allen:2004vz,Cartier:2003jz} and a conservation of the curvature perturbation on superhorizon scales during the expanding phase~\cite{Peter:2002cn,Kumar:2013koa,Battarra:2014tga}. The conservation of $\zeta_k$ on superhorizon scales can be viewed as well as a consequence of the local energy conservation which is valid for any relativistic gravitational theory~\cite{Lyth:2004gb,Wands:2000dp}. In view of these considerations, the curvature 
power spectrum at horizon crossing time during the RD era will be the same as
the curvature power spectrum at horizon exiting time during the initial $F(R)$ gravity
phase between the bounce and the RD era, namely
\beq
\mathcal{P}_{\zeta}\left[k,t_\mathrm{HC}(k,\alpha)\right] = \mathcal{P}_{\zeta}\left[k,t_\mathrm{exit}(k,\alpha)\right],
\eeq
where $t_\mathrm{HC}(k,\alpha)$ is given by (\ref{eq:t_HC}) and 
$t_\mathrm{exit}(k,\alpha)=\frac{k}{2\alpha}$. Finally, we can then use 
$\mathcal{P}_{\zeta}\left[k,t_\mathrm{HC}(k,\alpha)\right]$ and proceed to 
the calculation of the PBH abundance at horizon crossing time during the RD era, which is 
considered to be the PBH formation time.

\subsection{The scales involved}

 Regarding the relevant scales for the problem at hand, here we consider modes whose first horizon crossing time, i.e. when the modes exit the horizon, occurs before the RD era, that is $t_\mathrm{exit}<t_\mathrm{RD}$. Thus, accounting for the fact that $t_\mathrm{exit}(k,\alpha)=\frac{k}{2\alpha}$ and $t_\mathrm{RD}=1/\sqrt{\alpha}$, one can trivially find an upper bound on the comoving scale $k$ reading as
\beq\label{eq:k_bounds}
k<2\sqrt{\alpha}.
\eeq
This upper bound on $k$ is equivalent with a minimum PBH mass. In particular, considering the fact that the PBH mass is roughly the mass within the cosmological horizon at horizon crossing time during the RD era, one can trivially find that
{\bf {
\beq\label{eq:M_bounds}
M > \frac{2\pi \Mp^{2}}{\sqrt{\alpha}}
\eeq
}}

\section{The PBH Formation Formalism}\label{sec:PBH}

In this section we present  a general formalism for the computation of the 
mass function of PBHs formed due to the collapse of enhanced cosmological 
perturbations once they reenter the cosmological horizon. Basically, this 
happens when the energy density contrast of the collapsing overdensity region, 
or the respective comoving curvature perturbation, becomes greater than a 
critical threshold $\delta_\mathrm{c}$ or $\zeta_\mathrm{c}$. In the following, 
we firstly describe how the comoving curvature perturbation is connected to 
the energy density contrast, extracting  the non-linear relation between 
them, and then we proceed by presenting the formalism for the computation of 
the PBH mass function and the PBH abundance within the context of peak 
theory~\cite{Bardeen:1985tr}. At this point, it is important to highlight that we study PBH formation during the standard RD era described by general relativity. Therefore, the use of the peak theory formalism, developed within GR, for the computation of the PBH abundance is absolutely legitimate within our work. 

\subsection{From the comoving curvature perturbation to the energy density 
contrast}

Assuming spherical symmetry on superhorizon scales~\footnote{In principle, one could expect non spherical superhorizon perturbations due to the presence of an exotic equation of state with $w<-1$ after the bounce. In particular, the authors of~\cite{Yoo:2020lmg}, starting from spheroidal superhorizon perturbations and studying the role of non sphericities on the PBH threshold in the case of PBH formation during an RD era, found that their effect is negligibly small. Thus, as a first approximation, we will assume spherical symmetry on superhorizon scales as it is normally assumed in the literature.  However, in order to fully assess the effect of non sphericities on PBH formation due to the presence of a preceding exotic phase with a negative $w$ before RD era, one should perform high-cost numerical simulations which go beyond the scope of this work.}, the local region of the universe describing the aforementioned collapsing cosmological perturbations is described by the following asymptotic form of the metric 
\beq
\mathrm{d}s^2 = -\mathrm{d}t^2 + a^2(t)e^{\zeta(r)}\left[\mathrm{d}r^2 + r^2\mathrm{d}\Omega^2\right],
\eeq
where $a(t)$ is the scale factor and $\zeta(r)$ is the comoving curvature 
perturbation which is conserved on superhorizon scales.  In this regime one can 
perform a gradient expansion approximation, where all the hydrodynamic and 
metric quantities are nearly homogeneous, and their perturbations are small 
deviations away from their background values~\cite{Shibata_1999,Salopek:1990jq,Wands:2000dp,Lyth:2004gb}. In this approximation, the energy density 
perturbation profile is related to the comoving curvature perturbation through 
the following expression~\cite{Harada:2015yda,Yoo:2018kvb,Musco:2018rwt}:
\begin{align}\label{eq:zeta_vs_delta:non_linear}
\frac{\delta\rho}{\rho_\mathrm{b}} &\equiv\frac{\rho(r,t)-\rho_{\mathrm{b}}(t)}{\rho_{\mathrm{b}}(t)} \nonumber \\ & = -\left(\frac{1}{aH}\right)^2\frac{4(1+w)}{5+3w}e^{-5\zeta(r)/2}\nabla^2e^{\zeta(r)/2},
\end{align}
where  $w$ is the total equation-of-state parameter defined as the ratio 
between the total pressure $p$ and the total energy density $\rho$, i.e. 
$w\equiv p/\rho$.
In the linear regime, where $\zeta\ll 1$, the above expression is reduced to
\begin{align}\label{eq:zeta_vs_delta:linear}
\frac{\delta\rho}{\rho_\mathrm{b}} & \simeq -\frac{1}{a^2H^2}\frac{2(1+w)}{5+3w}\nabla^2\zeta(r) \nonumber \\ &  \Longrightarrow \delta_k =  -\frac{k^2}{a^2H^2}\frac{2(1+w)}{5+3w}\zeta_k.
\end{align}
Note that the last expression is obtained by Fourier 
transforming the energy density contrast $\delta$ and the curvature perturbation $\zeta$.

From the above  form we can see that there is a  one-to-one relation 
between the comoving curvature perturbation and the energy density contrast. 
Thus, if the curvature perturbation is a Gaussian variable then the same is 
true for the density contrast within the linear regime described by 
(\ref{eq:zeta_vs_delta:linear}). However, the amplitude of the critical 
threshold $\delta_\mathrm{c}$ or $\zeta_\mathrm{c}$ is in general non-linear, 
and as a consequence one should consider the full non-linear expression between 
$\zeta$ and $\delta$, namely (\ref{eq:zeta_vs_delta:non_linear}).

Here it is very important to stress that within the context of bouncing cosmological scenarios one expects in general the presence of non Gaussianities with an amplitude larger than the one predicted in simple inflationary setups~\cite{Quintin:2015rta,Gao:2014eaa}. In particular, for our case for perturbations whose first horizon crossing is before the onset of the RD era, the curvature perturbation $\zeta$ will become super-horizon during the intermediate exotic contracting phase with $w=-2/3$ possibly developing non-Gaussianity and eventually becoming highly non linear. After the onset of the RD era, due to the conservation of $\zeta$ in the expanding phase, it will remain constant. In view of these considerations we assume that the curvature perturbation field remains Gaussian and linear (to avoid breaking of perturbation theory) during the intermediate phase which connects the bounce with the RD era~\cite{Cai:2009fn}.

At this point, we should also highlight the fact that the use of $\zeta$ for the 
computation of the PBH abundance vastly overestimates the number of PBHs, since 
scales larger than the PBH scale, which are unobservable, are not properly removed when the 
PBH distribution is smoothed~\cite{Young:2014ana}. Therefore, one should 
instead use the energy density contrast,  given the fact that with this 
prescription the superhorizon scales are naturally damped by $k^2$, as it can be seen by 
(\ref{eq:zeta_vs_delta:non_linear}).

From a mathematical point of view, by performing a coordinate transformation on superhorizon scales, one can always shift the comoving curvature perturbation by an arbitrary constant, making the calculation of the PBH abundance not physical. On the other hand, if the density contrast is adopted instead, a dependence on spatial derivatives of the curvature perturbation is obtained as it can be seen by \Eq{eq:zeta_vs_delta:linear}, making the problem physical. This is another way to see that the choice to work with $\delta$ instead of $\zeta$ for the computation of the PBH abundance is the correct one.

Consequently, smoothing the energy density 
contrast with a Gaussian window function over scales smaller than the horizon scale and using (\ref{eq:zeta_vs_delta:non_linear}), we can straightforwardly 
find that  the smoothed energy density contrast is related to the comoving 
curvature perturbation in radiation era, where $w=1/3$, as~\cite{Young:2019yug}
\beq
\delta_\mathrm{m} = -\frac{2}{3}r_\mathrm{m}\zeta^\prime(r_\mathrm{m})\left[2+r_\mathrm{m}\zeta^\prime(r_\mathrm{m})\right].
\eeq
The scale $r_\mathrm{m}$ is the comoving scale of  the collapsing overdensity, 
which can be found by maximizing the compaction function $\mathcal{C}$ defined 
as~\cite{Musco:2018rwt}
\beq
\mathcal{C}(r,t) \equiv 2\frac{M(r,t)-M_\mathrm{b}(r,t)}{R(r,t)},
\eeq
where $R(r,t)$ is the areal radius, $M(r,t)$ is the Misner-Sharp 
mass~\cite{Misner:1964je, Hayward:1994bu} within a sphere of a radius $R$, and 
$M_\mathrm{b}=4\pi R^3(r,t)/3$ is the background mass with respect to a FLRW metric. Finally, by maximizing 
the compaction function, namely $\mathcal{C}^\prime(r_\mathrm{m})=0$, the 
$r_\mathrm{m}$ scale will be given by the solution of the following equation: 
\beq
\zeta^\prime(r_\mathrm{m}) + r_\mathrm{m}\zeta(r_\mathrm{m}) = 0.
\eeq

Now, given  the fact that $\zeta$ is assumed to have a Gaussian distribution, 
its derivative will have a Gaussian distribution too. Hence, we can 
identify a linear Gaussian variable 
$\delta_l=-\frac{4}{3}r_\mathrm{m}\zeta^\prime(r_\mathrm{m})$ with a probability 
distribution function (PDF) given by
\beq
P(\delta_l) = \frac{1}{\sqrt{2\pi\sigma}}e^{-\frac{\delta^2_l}{2\sigma^2}},
\eeq
where $\sigma$ is the smoothed variance of $\delta_l$  written as
\begin{align}\label{eq:sigma}
\sigma^2 & \equiv \langle \delta^2_l\rangle = \int_0^\infty\frac{\mathrm{d}k}{k}\mathcal{P}_{\delta_l}(k,R) \nonumber \\ & = \frac{16}{81}\int_0^\infty\frac{\mathrm{d}k}{k}(kR)^4 \tilde{W}^2(k,R)\mathcal{P}_\zeta(k).
\end{align}
The function $\tilde{W}(k,R)$ is the Fourier transformation of a Gaussian 
window function~\footnote{As regards the choice of the window function and its effect on the calculation of the PBH abundance see~\cite{Ando:2018qdb,Young:2019osy}.} and reads as
\beq
\tilde{W}(k,R) = e^{-k^2R^2/2},
\eeq

Finally, the smoothed energy density contrast is related with the linear Gaussian energy density contrast through the following expression~\cite{DeLuca:2019qsy,Young:2019yug}:
\beq\label{eq:delta_m_smoothed}
\delta_\mathrm{m} = \delta_l - \frac{3}{8}\delta^2_l.
\eeq

\subsection{The PBH mass function within peak 
theory}\label{eq:PBH_abundance_peak_theory}

In order to extract the mass function of PBHs which form due to the 
gravitational collapse of non-Gaussian energy density perturbations, we work 
  with the Gaussian component of the smoothed non-Gaussian energy density 
contrast denoted as $\delta_l$. Regarding the critical threshold of the linear Gaussian 
component, this can be found by solving \Eq{eq:delta_m_smoothed} for $\delta_l$ 
with $\delta_\mathrm{m} = \delta_\mathrm{c}$. Hence, we   find that
\beq
\delta_{\mathrm{c},l\pm} = \frac{4}{3}\left(1\pm 
\sqrt{\frac{2-3\delta_\mathrm{c}}{2}}\right).
\eeq
From the above expression we acquire a critical threshold for $\delta_l$. As 
explained in~\cite{Young:2019yug}, only $\delta_{\mathrm{c},l-}$ corresponds to 
a physical solution, and since the argument of the square root should 
be positive we require  $\delta_\mathrm{c}<2/3$. In summary, we find that  
the physical  range of $\delta_l$ is $\delta_{\mathrm{c},l-}<\delta_l<4/3$.

Regarding the PBH mass, it should be of the order of the horizon mass at PBH formation time, which is considered as the horizon crossing time. More precisely, the PBH mass spectrum, as it has been shown in~\cite{Niemeyer:1997mt,Niemeyer:1999ak,Musco:2008hv,Musco:2012au}, should follow a critical collapse scaling law which can be recast as
\beq\label{eq:PBH_mass_scaling_law}
M_\mathrm{PBH} = M_\mathrm{H}\mathcal{K}(\delta-\delta_\mathrm{c})^\gamma,
\eeq
where $M_\mathrm{H}$ is the mass within the cosmological horizon at horizon 
crossing time, and $\gamma$ is the critical exponent which depends on the 
equation-of-state   parameter at the time of PBH formation and for radiation 
it is $\gamma\simeq 0.36$. The parameter $\mathcal{K}$ is a 
parameter that depends on the equation-of-state parameter and on the particular 
shape of the collapsing overdensity region. In the following we consider a 
representative value of $\mathcal{K}\simeq 4$. 

Concerning now the value of the PBH formation threshold $\delta_\mathrm{c}$, its value should vary roughly within the range $0.4\lesssim \delta_\mathrm{c}\lesssim 0.6$ depending on the shape of the curvature power spectrum $\mathcal{P}_\zeta(k)$. Following the procedure developed in~\cite{Musco:2020jjb} we found that for the values of $\alpha$ studied here, namely for $\alpha\in [10^{-24}\Mp^2\leq\alpha\leq 10^{-14}\Mp^2]$, $\delta_\mathrm{c}\simeq 0.5898$ independently of the value of $\alpha$. This is somehow expected since as it can be seen from \Fig{fig:Pdelta_vs_alpha} the shape of $\mathcal{P}_\zeta(k)$ slightly changes with respect to $\alpha$. In particular, as one varies $\alpha$, we observe a change in terms of the overall amplitude of $\mathcal{P}_\zeta(k)$ and not in terms of its shape.

Thus, working with the Gaussian linear component of the  energy density contrast, we can calculate the PBH abundance in the context of peak theory, 
where the density of sufficiently rare and large peaks for a random Gaussian 
density field in spherical symmetry is given by~\cite{Bardeen:1985tr} 
\beq\label{eq:peak_density}
\mathcal{N}(\nu) = \frac{\mu^3}{4\pi^2}\frac{\nu^3}{\sigma^3}e^{-\nu^2/2}.
\eeq
In this expression,  $\nu \equiv \delta/\sigma$ and $\sigma$ is given by 
(\ref{eq:sigma}), while the parameter $\mu$ is the first moment of the 
smoothed power spectrum given by 
\begin{align}
\mu^2 & = 
\int_0^\infty\frac{\mathrm{d}k}{k}\mathcal{P}_{\delta_l}(k,R)\left(\frac{k}{aH}
\right)^2 = \nonumber \\ & \frac{16}{81}\int_0^\infty\frac{\mathrm{d}k}{k}(kR)^4 
\tilde{W}^2(k,R)\mathcal{P}_\zeta(k)\left(\frac{k}{aH}\right)^2.
\end{align}

Finally, the fraction $\beta_\nu$  of the energy of the universe at a peak of a 
given height $\nu$, which collapses to form a PBH, will be given by 
\beq\label{eq:peak_density}
\beta_\nu = \frac{M_\mathrm{PBH}(\nu)}{M_\mathrm{H}}\mathcal{N}(\nu)\Theta(\nu - 
\nu_\mathrm{c}) 
\eeq
and the total energy fraction of the universe contained in PBHs of mass $M$ can be recast as 
\beq\label{eq:beta_peak_theory}
\beta(M) = \int_{\nu_\mathrm{c-}}^{\frac{4}{3\sigma}}\mathrm{d}\nu\frac{\mathcal{K}}{4\pi^2}\left(\nu\sigma - \frac{3}{8}\nu^2\sigma^2 - \delta_{\mathrm{c}}\right)^\gamma \left(\frac{\mu}{\sigma}\right)^3\nu^3e^{-\nu^2/2},
\eeq
where $\nu_\mathrm{c-} = \delta_{\mathrm{c},l}/\sigma$. Lastly, the overall PBH 
abundance, defined as $\Omega_\mathrm{PBH}\equiv 
\frac{\rho_\mathrm{PBH}}{\rho_\mathrm{tot}}$, where $\rho_\mathrm{tot}$ is the 
total energy density of the universe, will be the integrated PBH mass function. Thus, at  time $t$ during the RD era, $\Omega_\mathrm{PBH}$ will be recast as
\beq\label{eq:Omega_PBH_t}
\Omega_\mathrm{PBH} (t) = \int_{M_\mathrm{min}}^{M_\mathrm{max}}  \left(\frac{M_\mathrm{H}(t)}{M}\right)^{1/2}\beta(M)\mathrm{d}\ln M,
\eeq
where $M_\mathrm{H}(t)$ is the mass within the cosmological horizon at time $t$. Note that in \Eq{eq:Omega_PBH_t} we have accounted for the fact that during the RD era $M_\mathrm{H}\sim a^{2}$.

\section{Results}\label{sec:results}

In  the previous sections we extracted  the curvature power spectrum and we
presented the mathematical setup through which one can calculate the PBH mass 
function and abundance during the standard RD era which follows the exotic $F(R)$ gravity phase close to the bounce. Thus, in this section we present the main results 
of our work. Initially, we study the behaviour of the curvature power 
spectrum by varying the parameters of the problem at hand, namely the bouncing parameter $\alpha$. Then, we 
compute numerically the PBH mass function and we show how it varies by changing $\alpha$.
Finally, by demanding that GWs induced from PBH Poisson fluctuations during an early PBH dominated era before BBN are not overproduced, we set constraints on $\alpha$.

\subsection{The curvature power spectrum}\label{sec:Pdelta}

Given the fact that the scales collapsing to PBHs are initially super-Hubble 
before crossing the  Hubble radius and collapse to PBHs, we perform a Taylor 
expansion of the comoving curvature perturbation   (\ref{eq:zeta_k}) on 
super-Hubble scales, i.e. when $k\ll aH$. By keeping terms up to 
$\mathcal{O}\left[\left(\frac{k}{aH}\right)^{3/2}\right]$ we obtain that

\begin{align}\label{eq:zeta_k_superH}
& \zeta_{k,k\ll aH}  \simeq  \kappa e^{3\alpha t^2}\sqrt{\frac{3}{t(12\alpha - 
C)}}\left[\frac{k}{a(t)H(t)}\right]^{-1/2}  \nonumber \\ &
-i\frac{\kappa\alpha t }{\sqrt{t(12\alpha - C)}}\left[e^{3\alpha t^2}\sqrt{\pi} - 
2H\left(-1,t\sqrt{3\alpha}\right)\right]\sqrt{\frac{k}{a(t)H(t)}}  \nonumber \\ 
&  -\frac{\kappa\alpha t^{3/2}}{\sqrt{3(12\alpha -  
C}}{}_1F^{(1,0,0)}_1\left(\frac{1}{2},\frac{1}{2},3\alpha t^2\right)\left[\frac{k}{a(t)H(t)}
\right]^{3/2},
\end{align}
where $_1F^{(1,0,0)}_1(x,y,z)$ stands for the derivative of the Kummer confluent hypergeometric function with respect to its first argument.

Therefore, inserting this expression in \Eq{eq:P_zeta_close_to_the_bounce} and following the procedure described in \Sec{sec:PBH}, we can calculate the curvature power spectrum $\mathcal{P}_\zeta(k)$ at horizon crossing time by fixing the bouncing parameter $\alpha$ and the integration 
constant $C$. As it was checked numerically, $\mathcal{P}_\zeta(k)$ is 
independent on the value of $C$ and in the following we will fix its value to $C=0.1\alpha$. In the following, we will use the above expression for $\zeta_{k,k\ll aH} $ when computing the comoving curvature perturbation and subsequently the matter power spectrum $\mathcal{P}_\delta(k)$ following the procedure described in \Sec{sec:PBH}. As it was confirmed numerically the curvature power spectrum $\mathcal{P}_\zeta(k)$ computed using \Eq{eq:zeta_k_superH} matches quite well the exact $\mathcal{P}_\zeta(k)$  all along the $k$ range. 

\begin{figure}[h!]
\begin{center}
\includegraphics[width=0.495\textwidth]{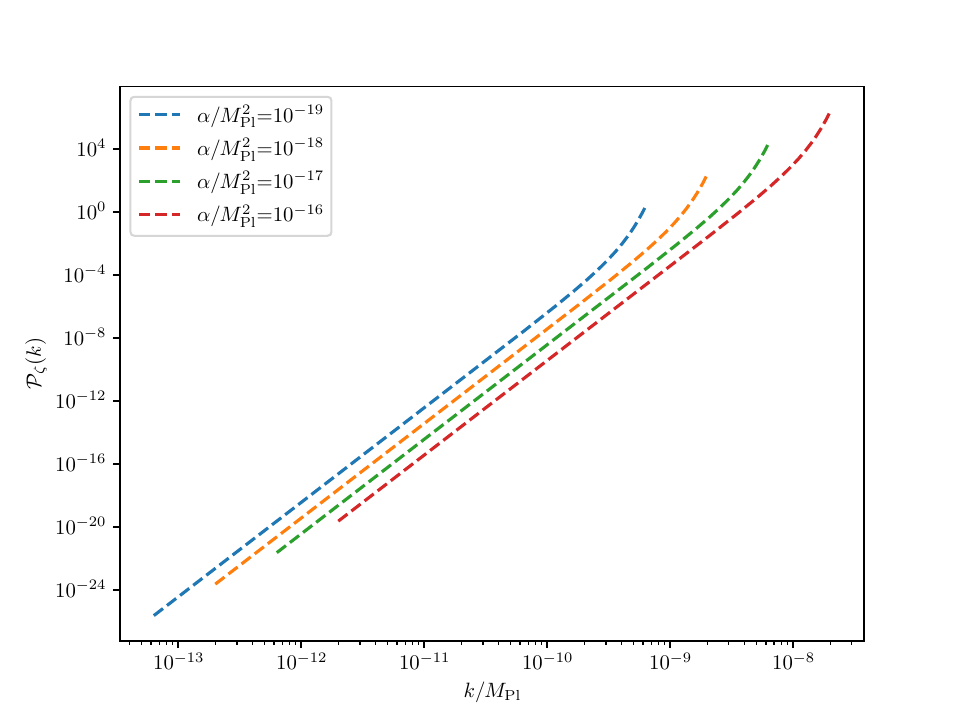}
\caption{{\it{The curvature power spectrum versus $k$ for different 
values of $\alpha$.}}}
\label{fig:Pdelta_vs_alpha}
\end{center}
\end{figure}

In Fig. \ref{fig:Pdelta_vs_alpha}, we depict the curvature
power spectrum $\mathcal{P}_\zeta(k)$ [\Eq{eq:P_zeta_close_to_the_bounce}] on superhorizon scales, for different values of $\alpha$ and for $C=0.1\alpha$. As we can see, the power spectrum increases by increasing 
the value of $\alpha$. This behaviour 
can be understood if one sees how the maximum allowed value of $k$, which 
corresponds to the lowest scale of the problem at hand, varies with $\alpha$. In particular, as we can see from \Eq{eq:k_bounds}, the value of $k_\mathrm{max}$ increases with an increase of $\alpha$, hence
the power spectrum shifts to higher values of $k$, i.e. to smaller scales. 
Consequently, as approaching smaller and smaller scales one starts to probe the 
granularity of the energy density field, entering in this way the non linear 
regime where $\mathcal{P}_\mathrm{\zeta}(k)\gg 1$. Hence, one can clearly 
understand the tendency of the power spectrum to increase with increasing 
$\alpha$, given the fact that it probes smaller scales which become non-linear.

In order to avoid the presence of non linearities, one could abruptly cut the curvature power spectrum at values smaller than unity in order to ensure the validity of the linear perturbative regime. However, given the fact that PBH formation is a non-linear process since it takes place in overdensity regions where $\delta>\delta_\mathrm{c}\sim \mathcal{O}(1)$ the introduction of an abrupt cutoff would dramatically decrease the PBH abundance to values orders of magnitude smaller than its real value. The correct way to remove these non-linear scales is actually through the introduction of the non-linear transfer function which has not yet been extracted and requires high cost $N$ body simulations which go beyond the scope of this work~\cite{Young:2019osy}. Consequently, as it is standardly adopted within the context of the PBH literature, these small non-linear scales are naturally smoothed out when computing the PBH mass function through the use of a window function introduced in \Sec{sec:PBH}.

\subsection{The PBH mass function}\label{sec:beta}

\begin{figure*}[t!]
\begin{center}
\includegraphics[width=0.495\textwidth]{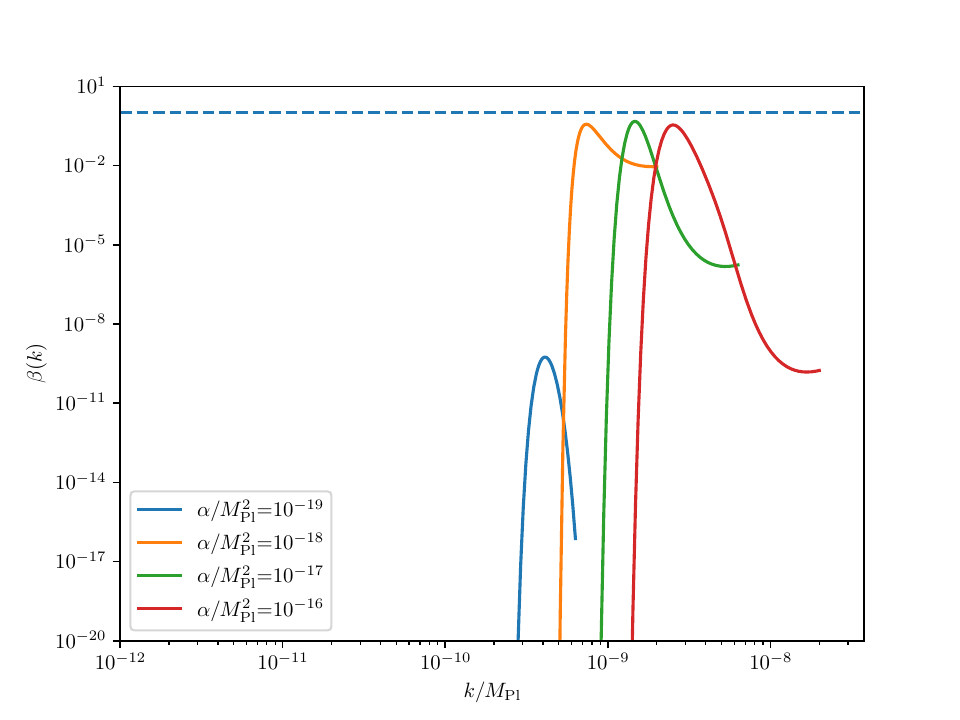}
\includegraphics[width=0.495\textwidth]{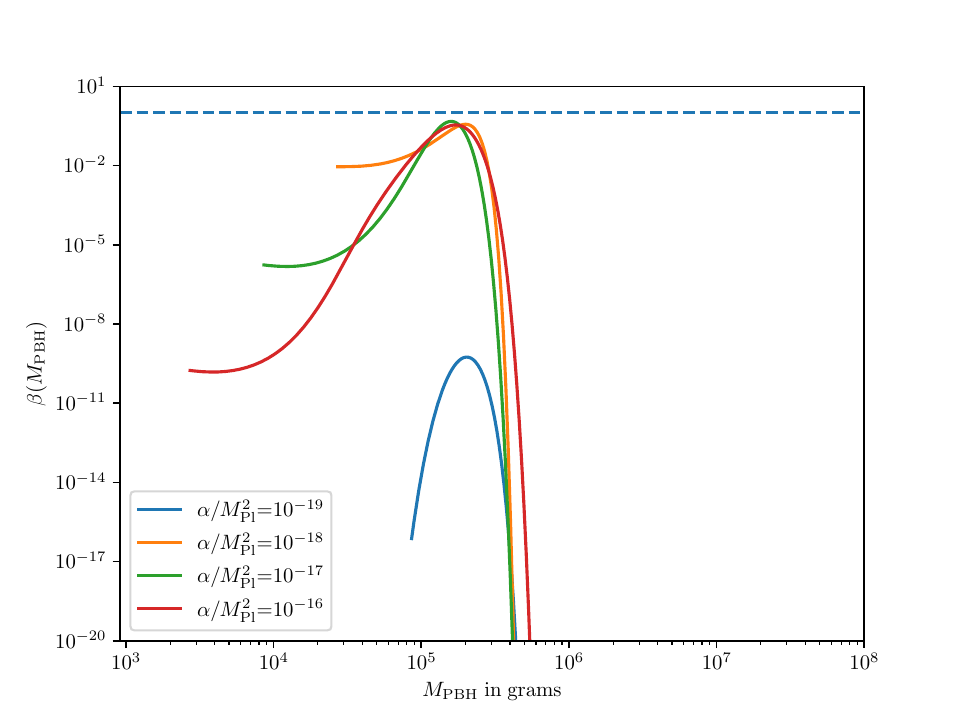}
\caption{{\it{Left panel: The PBH mass function $\beta(k)$ as a function of the comoving number $k$ for different values of the $F(R)$ bouncing parameter $\alpha$. Right panel: The PBH mass function $\beta(M)$ as a function of the PBH mass $M_\mathrm{PBH}$ for different values of the $F(R)$ bouncing parameter $\alpha$. The blue dashed horizontal line stands for $\beta(M) = 1$.}}}
\label{fig:beta_vs_alpha}
\end{center}
\end{figure*}

Since we have  extracted above the curvature power spectra for different values of $\alpha$, we proceed to the 
calculation of the PBH mass function within peak theory. In particular, we follow the 
mathematical formalism presented in \Sec{eq:PBH_abundance_peak_theory}, accounting for the non-linear relation between $\delta$ and $\zeta$ as well as the critical collapse law for the PBH masses. Below, we show how the PBH 
mass function changes by varying the parameter $\alpha$. As a first general comment, one may notice from \Fig{fig:beta_vs_alpha} that we are met with an extended PBH mass distribution as it can be expected if one sees \Fig{fig:Pdelta_vs_alpha} where $\mathcal{P}_\zeta(k)$ is not peaked but instead varies over a wide range of comoving scales $k$.

In the left panel of \Fig{fig:beta_vs_alpha}, we show how the PBH mass function changes with respect to the comoving scale $k$ for different values of the parameter $\alpha$. In particular, the mass function increases its overall amplitude as one increases the value of the parameter $\alpha$, a behavior which is kind of expected since as explained in \Sec{sec:Pdelta} by increasing $\alpha$ one starts to probe more and more smaller scales which become non-linear and can easily collapse to PBHs. 

Interestingly, one can also notice that for values of $\alpha$ more or less larger than $10^{-19}\Mp^2$, the peak of the mass function saturates at a value close to $0.1$ independently of the value of $\alpha$. This behavior can be explained if one sees \Fig{fig:Pdelta_vs_alpha} where we see that for $\alpha>10^{-19}\Mp^2$, the curvature power spectrum enters gradually as we increase the value of $\alpha$ deep into the non perturbative regime where $P_\zeta(k)\gg 1$. Consequently, due to the effect of smoothing these enhanced perturbation modes do not contribute to the increase of the mass function as we go to high $k$ values. On the contrary, the overall effect of smoothing is to make the maximum amplitude of $\beta$ to saturate for $\alpha>10^{-19}\Mp^2$.

One can also infer a shift of the position of the peak of $\beta(k)$ towards the smaller scales, namely large $k$ values, a behavior which can be explained from the fact that $k_\mathrm{max}\sim \sqrt{\alpha}$ [See \Eq{eq:k_bounds} ]. 

Additionally, we witness as well a slight increase on the large $k$ region. This slight increase is due to the fact that in the high $k$ region where $\delta$ is very large, the PBH mass function \eqref{eq:beta_peak_theory} scales as $\beta(M)\propto 1/\sigma^6$ with $\sigma^2$ being suppressed on the very small PBH scales due to the effect of smoothing which becomes very important on these scales. As a consequence, at a scale around $k_*\sim k_\mathrm{max}/4$ all $\beta(k)$ curves start to slightly increase as one probes smaller scale modes $k$. [See the discussion in  \App{app:sigma2_mu2}.] 

In the right panel of \Fig{fig:beta_vs_alpha}, we show how the $\beta$ function changes with respect to the PBH mass by varying the parameter $\alpha$. The observed behavior is similar as in the left panel of \Fig{fig:beta_vs_alpha} with the only difference that now the position of the peak of $\beta(M)$ is more or less constant, independent of the value of $\alpha$. This can be understood if we see how the PBH mass scales with $\alpha$ and $k$. In particular, by defining the PBH mass being roughly equal to the mass within the horizon at horizon crossing time during the RD era one obtains that
\beq \label{eq:M_PBH_vs_alpha_k}
M_\mathrm{PBH} \simeq M_\mathrm{H} = \frac{4\pi\Mp^2}{H} = \frac{8\pi\Mp^2\sqrt{\alpha}}{k^2},
\eeq
where in the last step we used \Eq{eq:t_HC} as well as the fact that during the RD era $H = 1/(2t)$. 
Thus, despite the fact that as one increases the value of $\alpha$ the position of the peak of the $\beta$ function shifts to higher values of $k$, i.e. smaller scales [See left panel of \Fig{fig:beta_vs_alpha}] when one plots $\beta$ in terms of $M_\mathrm{PBH}$ the position of the peak of $\beta$ will shift to larger masses, since $M_\mathrm{PBH}\sim \sqrt{\alpha}/k^2$ as it can be seen by \Eq{eq:M_PBH_vs_alpha_k}. At the end, the overall effect is that the position of the peak of the function $\beta(M_\mathrm{PBH})$ is more or less constant independently of the value of $\alpha$.

At this point, it is useful to stress that the PBH masses produced substantially by the $F(R)$ gravity bouncing model studied here are very small, namely less than $10^9\mathrm{g}$, evaporating very quickly before the BBN time. One question one could ask is if with this bouncing model one can produce higher PBH masses, close to the solar mass as the ones probed by LIGO/VIRGO gravitational-wave detectors. To give an order of magnitude of the value that the $F(R)$ gravity parameter $\alpha$ should have in order to produce PBH masses of the order of $1M_\odot$ we can simply set in \Eq{eq:M_PBH_vs_alpha_k} $M_\mathrm{PBH}=1M_\odot$ and the comoving value $k$ equal to its maximum value, namely $k=k_\mathrm{max}=2\sqrt{\alpha}$. At the end, one gets straightforwardly  that 
\beq
M_\mathrm{PBH}>M_\odot\Leftrightarrow\alpha<4\times 10^{-72}\Mp^2.
\eeq
For such very small values of $\alpha$ the PBH mass function is dramatically suppressed as one may speculate by looking at the decreasing tendency of $\beta$ by decreasing the value of the parameter $\alpha$ in \Fig{fig:beta_vs_alpha}.

\subsection{Constraining $\alpha$}

\begin{figure}[h!]
\begin{center}
\includegraphics[width=0.495\textwidth]{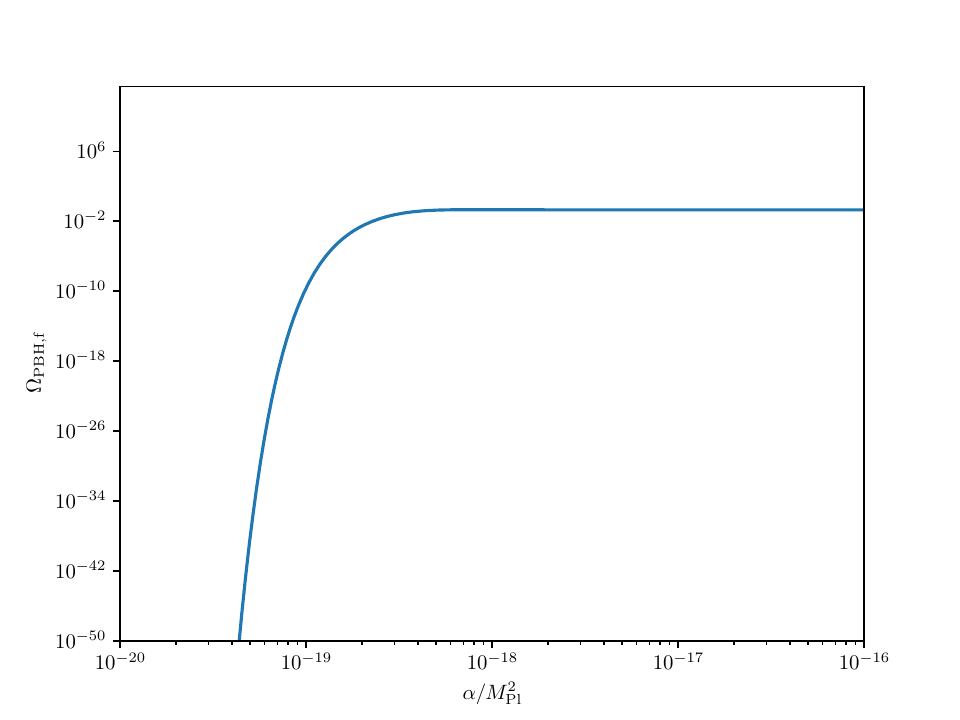}
\caption{{\it{The PBH abundance at formation time $\Omega_\mathrm{PBH,f}$ as a function of the $F(R)$ bouncing parameter $\alpha$.}}}
\label{fig:Omega_PBH_f}
\end{center}
\end{figure}

We can now proceed to  perform a full parameter-space analysis by calculating  the PBH abundance at formation time $\Omega_\mathrm{PBH,f}$, for a 
wide range of values of the $F(R)$ parameter $\alpha$. 
In \Fig{fig:Omega_PBH_f} we show how $\Omega_\mathrm{PBH,f}$ varies 
as a function of the bouncing parameter $\alpha$. In particular, we find that as $\alpha$ increases, the PBH abundance increases as well, as it can been speculated from \Fig{fig:beta_vs_alpha}. This behavior can be explained from the fact that as $\alpha$ increases the curvature power spectrum shifts to smaller and smaller scales widening in this way the range of modes $k$ which can potentially collapse to PBHs, hence enhancing the PBH mass function. Interestingly, we find that for values $\alpha\geq 10^{-19}\Mp^2$, $\Omega_\mathrm{PBH,f}$ saturates to a plateau which is related with the saturation of the amplitude of the PBH mass function due to the effect of smoothing becoming more and more important as $\alpha$ increases [See the discussion in \Sec{sec:beta}.].

At the end, accounting for the fact that the masses of the formed PBHs are so
small they evaporate very quickly after their formation. Consequently, the only natural condition which needs to be fulfilled so as to set constraints on the parameter $\alpha$ is that $\Omega_\mathrm{PBH,f}<1$. However, as recently noted in~\cite{Papanikolaou:2020qtd} such small PBHs evaporating before BBN can dominate the energy budget of the Universe and induce at second order in cosmological perturbation theory a GW background which can be detectable by future GW experiments. Requiring therefore that GWs are not overproduced during this early PBH dominated era, one  can set constraints on the parameters of the PBH production mechanism and in our case the $F(R)$ gravity parameter $\alpha$. For the case of monochromatic PBH distributions one can show that in order for the GWs not to be overproduced one should require that ~\cite{Papanikolaou:2020qtd} 
\beq\label{eq:Omega_PBH_f_bounds_monochromatic}
\Omega_\mathrm{PBH,f}<10^{-4}\left(10^9\mathrm{g}/M_\mathrm{PBH}\right)^{1/4}.
\eeq
In our case, we have a broad PBH mass spectrum but given the fact that the position of the peak of the maximum of the PBH mass function depends slightly on the value of the parameter $\alpha$ we can use as a first approximation \Eq{eq:Omega_PBH_f_bounds_monochromatic} in order to constrain the bouncing parameter $\alpha$. In order to be more precise, one should account for the full broad PBH mass distribution and compute the GW signal today accounting as well for the transition between the early PBH dominated era to the RD era~\cite{Papanikolaou:2022chm}, a study which goes beyond the scope of the present work and which we leave for a future project.

Thus, taking $M_\mathrm{PBH}\simeq 2\times 10^ 5\mathrm{g}$ which is more or less the PBH mass at the peak of the $\beta$ function one gets that $\Omega_\mathrm{PBH,f}<10^{-3}$. At the end, requiring this condition one finds numerically [See \Fig{fig:Omega_PBH_f}] that 
$\alpha$ should lie within the following range:
\beq
\alpha \leq 10^{-19}M^2_\mathrm{Pl}.
\eeq

This constraint can be translated to constraints on the energy scale at the onset of the HBB phase  $H_\mathrm{RD}$ given the fact that $t_\mathrm{RD}=1/\sqrt{\alpha}$ and $H_\mathrm{RD} = 1/(2t_\mathrm{RD})$. At the end, one can find that $H_\mathrm{RD}=\sqrt{\alpha}/2$ and should vary within the following range:
\beq
H_\mathrm{RD}\leq 10^{-10}M_\mathrm{Pl}.
\eeq

At this point, it is very important to stress that the energy scale at the onset of the RD era, given by $H_\mathrm{RD}$, can also be viewed as the lowest bound on the energy scale of the Universe at the bounce.

\section{Conclusions}\label{sec:conclusions}

The non-singular bouncing cosmological paradigm  is one of the most appealing 
alternatives to inflation. Since the bounce realization requires the violation 
of the null energy condition, it can be typically   implemented in the 
framework of modified gravity. On the other hand, the phenomenology of PBH 
physics, and the associated PBH 
abundance constraints which span a range of masses over more than $50$ orders 
of magnitude, has recently started to be investigated in detail, since it can 
be used in order to probe and extract constraints on the early-universe 
behavior. Hence, studying PBHs both at inflationary and bounce scenarios, could 
be helpful to constrain such scenarions and extract possible distinguishable features.

In this work, we focused on the bounce realization within $F(R)$ modified gravity
and we investigated the corresponding PBH phenomenology. By introducing an $F(R)$ gravity exotic phase close to the bounce compatible with a bouncing scale factor we studied its effect on the mass function of PBHs which form during the standard RD era described quite well within classical GR gravity. In particular, we calculated the curvature power spectrum at 
horizon crossing time, during the RD era, as a function of the  the bounce parameter
$\alpha$, which is actually the involved $F(R)$ gravity parameter.

Followingly, we calculated the PBH abundance in the 
context of peak theory, considering the non-linear relation between $\delta$ 
and $\zeta$ as well as the critical collapse law for the PBH masses. At the end, in 
\Fig{fig:beta_vs_alpha} we showed how the PBH mass 
function changes by varying the bouncing parameter $\alpha$.

Additionally, by making a full parameter-space analysis, in 
\Fig{fig:Omega_PBH_f} we gave the PBH abundance at formation time 
$\Omega_\mathrm{PBH,f}$ as a function of the bouncing parameter $\alpha$.
Interestingly enough, we found that in order to 
avoid GW overproduction from an early PBH domination era before BBN, $\alpha$ should lie 
within the range $\alpha \leq 10^{-19}M^2_\mathrm{Pl}$. This constraint can be transformed to a constraint on the energy scale at the onset of the HBB phase $H_\mathrm{RD}\sim \sqrt{\alpha}/2$ which can be recast as $H_\mathrm{RD}\leq 10^{-10}M_\mathrm{Pl}$.

We mention that the explored  parameter space can 
be further constrained by evolving the PBH abundance $\Omega_\mathrm{PBH}$ up to 
later times, and accounting for current observational constraints on 
$\Omega_\mathrm{PBH}$~\cite{Carr:2020gox}. Moreover, one can extract more stringent constraints 
by studying  additionally the scalar induced stochastic gravitational-wave 
background (SGWB) associated to the primordial curvature perturbations which 
gave rise to PBHs (see  \cite{Domenech:2021ztg} for a review), as well as 
the SGWB induced from PBH Poisson 
fluctuations~\cite{Papanikolaou:2020qtd,Papanikolaou:2021uhe, 
Papanikolaou:2022hkg,Papanikolaou:2022chm}.

Since PBH formation within bouncing cosmologies may serve as a novel 
tool to study alternative theories of gravity, one should perform a 
similar analysis in other modified gravity scenarios, and examine whether there 
are qualitative and quantitative differences amongst them. In particular, one can extend our formalism by accounting as well for the effect of modified gravity on the background and perturbation evolution during the period of PBH formation generalising in a sense the peak theory formalism and investigating the full gravitational collapse dynamics in modified gravity setups. Such a detailed 
investigation is beyond the scope of this paper and can be be performed elsewhere.

\begin{acknowledgments}
T.P. acknowledges financial support from the Foundation for Education
and European Culture in Greece. T.P. would like to thank as well the Laboratoire Astroparticule and Cosmologie, CNRS Université Paris Cité for kind hospitality as well as for giving him access to the computational cluster DANTE where part of the numerical computations of this paper were performed. The authors would like to acknowledge the contribution of the COST Action CA18108 ``Quantum Gravity Phenomenology in the multi-messenger approach''. 
\end{acknowledgments}
\begin{appendix}
\section{Investigating different bouncing scale factor parametrisations}\label{app:different_a_parametrisations}
Up to now, we have considered that the scale factor close to the bounce is 
parametrized by (\ref{eq:a_close_to_the_bounce}), by keeping terms up to 
quadratic order in $t$ in the Taylor   expansion for $a(t)$. Thus, a legitimate question to ask is how our results will change 
by changing the scale factor parametrisation near the bounce. In general, the 
scale factor near a non-singular bounce can be parameterized 
as~\cite{Odintsov:2020zct} 
\beq
a_\mathrm{b}(t) \simeq (1 + \alpha t^2)^n,
\eeq
where $n$ is a real number. In the following  we study the cases where $n=2$ 
and $n=3$ and we examine how the curvature power spectrum changes accordingly.

 1) $a(t)=(1+\alpha t^2)^2$:
 
 Using this parametrisation for the scale factor near the bounce, and solving 
\Eq{basic4} for $F(R)$ we find that 
\begin{eqnarray}
F_b(R(t)) && =\frac{1}{420}t^2\alpha^2\Big(99225+t^2 \alpha \Big(-814275 
+  \nonumber \\ &&  t^2 \alpha \Big(91875  + 
      t^2 \alpha \Big(15855 + 
         t^2 \alpha \Big(3360 + 
            t^2 \alpha \Big(245  \nonumber \\ &&  + 
               t^2 \alpha [75  + 
                  t^2 \alpha (-25 + t^2 
\alpha)]\Big)\Big)\Big)\Big)\Big)\Big) \nonumber \\ && + \frac{1}{\alpha^5}(105 + t^2 \alpha (525  +  t^2 \alpha(-1050 +  t^2 \alpha [350  \nonumber \\ && + t^2 \alpha (-35 + t^2 
\alpha)])))C +
\frac{1}{8 \sqrt{t^2 \alpha}}9 \pi t 
\alpha^{3/2}(105  \nonumber \\ && + 
t^2 \alpha (525 + t^2 \alpha (-1050 + 
t^2 \alpha [350  + t^2 \alpha (-35 \nonumber \\ && + t^2\alpha)]))) \cdot
\mathrm{Erfc}(t\sqrt{t \alpha}/\sqrt{2})\mathrm{Erfi}(t\sqrt{t 
\alpha}/\sqrt{2}) \nonumber \\ && + \frac98 \alpha 
(-e^{t^2 \alpha/2} \sqrt{2 \pi}t \sqrt{\alpha} 
(105 + 
    t^2 \alpha (-790 \nonumber \\ && + 
       t^2 \alpha[318 + t^2 \alpha (-34 + t^2 
\alpha)]))  +\pi (105 + 
   t^2 \alpha \nonumber \\ && (525 + 
      t^2 \alpha(-1050 + 
         t^2 \alpha [350 + t^2 \alpha (-35  + t^2 
\alpha) \nonumber \\ && ])))
\cdot\mathrm{Erfi}(t\sqrt{t \alpha}/\sqrt{2}))\mathrm{Erf}(t\sqrt{t \alpha}/\sqrt{2}) \nonumber \\ && -\frac{1}{840}e^{t^2 \alpha/2}t^{12} \alpha^7 (105 + 
   t^2 \alpha (-790 + 
      t^2 \alpha [318 \nonumber \\ && + t^2 \alpha (-34 + t^2 
\alpha)]))\mathrm{ExpIntE}(-\frac92, \frac{t^2 
\alpha}{2}),
\end{eqnarray}
where $\mathrm{ExpIntE}$ is the exponential integral function $E_n(z)$.

Similar to the previous case, keeping terms up to $\mathcal{O}(\alpha t^2)$ the 
expression for $z(t)$ becomes
\begin{equation}
    z(t)=U+Vt-Xt^2,
\end{equation}
where 
\beq
U =\frac{\sqrt{105}\alpha}{2\alpha\kappa\sqrt{\alpha^7/C}},
\eeq
\beq
V =  \frac{\sqrt{\alpha^7/C}(2592\pi\alpha^{12}-1225C^2)}{12\alpha^{12}\kappa\sqrt{210\pi}},
\eeq
\begin{align} 
X & =\frac{(\alpha^7/C)^{3/2}}{1890\alpha^{18}\kappa\pi\sqrt{105}}\bigl[419904\pi^2\alpha^{18}+45360\pi\alpha^{12}C \nonumber \\ & +1190700\pi\alpha^6C^2+42875C^3\bigr].
\end{align}

Thus, evaluating   the curvature perturbation near the bounce, at leading order 
in $t$, we obtain  
\begin{align}
\zeta_k & =C_1(k)H\bigg[\frac{k^2U^2}{2V^2+4UX-4U^2\alpha}, \nonumber \\ & \frac{-UV+tV^2+2 
tUX-2tU^2\alpha}{U\sqrt{V^2+2U(X-U\alpha)}}\bigg] \\  \nonumber 
& +C_2(k)~{}_1F_1\bigg\{-\frac{k^2U^2}{4[V^2+2U(X-U\alpha)]},\frac12, \nonumber \\ & 
\frac{[-tV^2+U(V-2tX)+2tU^2\alpha]^2}{U^2[V^2+2U(X-U\alpha)]}\bigg\}\nonumber .
\end{align}
The forms of $C_1(k)$ and $C_2(k)$ are determined using the initial conditions 
given in   (\ref{initial}) modified appropriately for the present case where $a(t)=(1+\alpha t^2)^2.$ 
 
\begin{figure*}[t!]
\begin{center}
\includegraphics[width=0.495\textwidth]{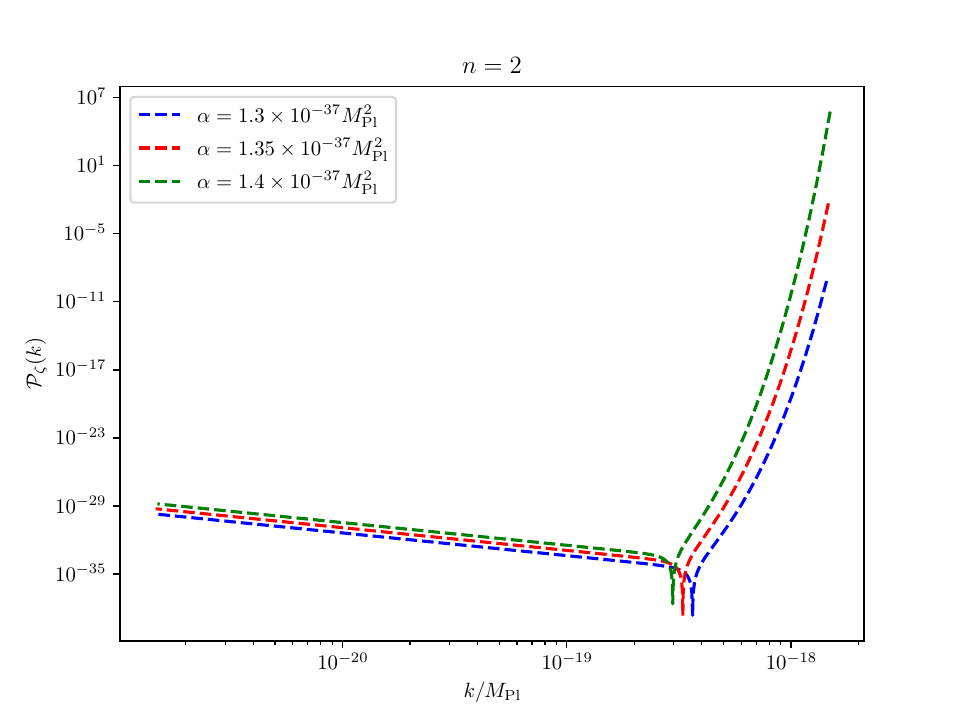}
\includegraphics[width=0.495\textwidth]{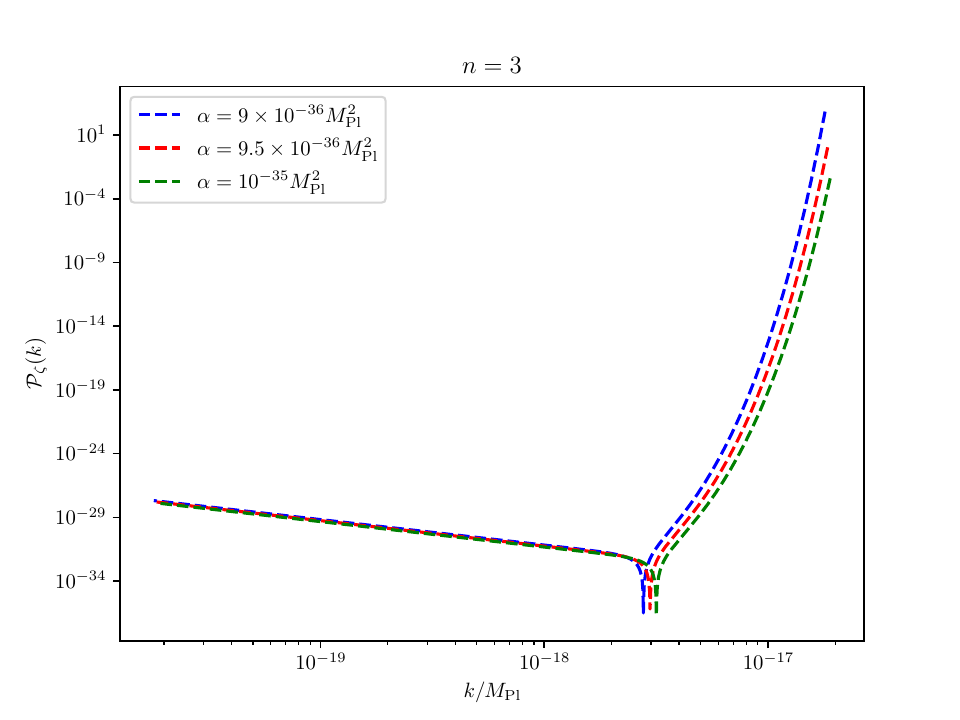}
\caption{{\it{The curvature power spectrum $\mathcal{P}_\zeta(k)$ for different values of $\alpha$ for $n=2$ and $n=3$.}}}
\label{fig:P_zeta_vs_alpha_n}
\end{center}
\end{figure*}

Below, we show the curvature power spectrum $\mathcal{P}_\zeta(k)=k^3|\zeta_k|^2/(2\pi^2)$ by varying the $F(R)$ bouncing parameter $\alpha$.
As one may notice from the left panel of \Fig{fig:P_zeta_vs_alpha_n} in the case where $n=2$, $\mathcal{P}_\zeta(k)$ becomes very sensitive with $\alpha$ with a general tendency to increase on small scales, i.e. large $k$ values, probing gradually the non linear regime. In addition, it is worth highlighting the fact that independently of the value of $\alpha$ $\mathcal{P}_\zeta(k)$ increases very abruptly to large values within less than one order of magnitude in $k$ signalling the fact that in contrast with the $n=1$, one is met with an almost monochromatic curvature power spectrum giving rise to PBHs.

2) $a(t)=(1+\alpha t^2)^3$:

With the same reasoning as before, the solution for $F(R)$ around the bounce 
reads as
\begin{eqnarray}
F_b(t)&&=6 
\alpha+324\alpha^2t^2+\Big[324t^4\alpha^3(-3+t^2\alpha) 
\nonumber \\ && -\frac{54}{288}e^{\frac{3t^2\alpha}{2}}t\alpha^2 
\left[1 +t^2\alpha(-8+3t^2\alpha)\right]C  \nonumber \\ && +9C 
\sqrt{6\pi}\alpha^{9/2}(1+9t^2\alpha[1+t^2\alpha(-3 \nonumber \\ && +t^2\alpha)]) \mathrm{Erfi}\left(\sqrt{\frac{3\alpha}{2}}t\right)\Big].
\end{eqnarray}
Once again, keeping up to $\mathcal{O}(\alpha t^2)$ terms in the scalar perturbation, 
we extract the form of $z(t)$ as
\begin{equation}
    z(t)=U+Vt-Xt^2,
\end{equation}
where 
\begin{eqnarray}
 U=&&\frac{\sqrt{6}}{\kappa}, \quad V=\frac{(-124416+\alpha 
C^2)}{24\sqrt{6}\kappa C},  \\  X=&&\frac{ \alpha(746496+ \alpha C^2) }{6912 
(\sqrt{6}\kappa)}.
\end{eqnarray}
The corresponding solution for the curvature perturbation, at leading order in 
$t$, is
\begin{align}
\zeta_k & =C_1(k)H\bigg[\frac{k^2U^2}{2V^2+4UX-4U^2\alpha}, \nonumber \\ & \frac{-UV+tV^2+2 
tUX-2tU^2\alpha}{U\sqrt{V^2+2U(X-U\alpha)}}\bigg] \\  \nonumber 
& +C_2(k)~{}_1F_1\bigg\{-\frac{k^2U^2}{4[V^2+2U(X-U\alpha)]},\frac12, \nonumber \\ & 
\frac{[-tV^2+U(V-2tX)+2tU^2\alpha]^2}{U^2[V^2+2U(X-U\alpha)]}\bigg\}\nonumber .
\end{align}

In the right panel of \Fig{fig:P_zeta_vs_alpha_n} we show again the curvature power spectrum for the $n=3$ case by varying the parameter $\alpha$. In particular, as in the $n=2$, one can notice a power spectrum $\mathcal{P}_\zeta(k)$ with an amplitude quite sensitive to the variation of the $F(R)$ bouncing parameter $\alpha$ and with a tendency to lead to a monochromatic PBH mass distribution in contrast with the $n=1$ case.

Consequently, one can 
argue that our results are nearly the same for $(1+\alpha t^2), \ (1+\alpha 
t^2)^2, \ (1+\alpha t^2)^3 $ and other values of $n$ in $(1+\alpha t^2)^n$ with $n>1$. In particular, in contrast with the $n=1$ case, we find a very sensitive behavior of the amplitude of $\mathcal{P}_\zeta(k)$ and a tendency of $\mathcal{P}_\zeta(k)$ to lead to a monochromatic PBH mass function.

Finally, one should comment on the order of masses produced within the parametrisations where $n>1$. In particular, as we can see fron \Fig{fig:P_zeta_vs_alpha_n} $k_\mathrm{max}\sim 10^{-18}\Mp$ and given the fact that $M_\mathrm{PBH}\propto \sqrt{\alpha}/k^2$ one gets that for $\alpha \sim 10^{-36}\Mp^2$ $M_\mathrm{PBH}\sim 10^{13}\mathrm{g}\sim 10^{-20}M_\odot $ many orders of magnitude larger than the order of PBH masses produced in the $n=1$ case but still quite small compared to the PBH masses detected by the LIGO-VIRGO detectors which are of the order of the solar mass.  

We mention here that  
other possible bouncing scale factor forms that have been studied in the 
literature  are $\cosh\left(1+\alpha t^2\right)$ and $e^{\alpha t^2}$. However, when 
expanded around $t$  their forms become similar to $(1+\alpha t^2)^n$, hence our above results become quite general, being valid for any parametrization of 
the scale factor giving rise to a bounce.

\section{The PBH mass function on small scales}\label{app:sigma2_mu2}
We show below the smoothed power spectra $\sigma^2$ and $\mu^2$ with respect to the comoving scale $k$ by varying the $F(R)$ gravity parameter $\alpha$. 

\begin{figure*}[t!]
\begin{center}
\includegraphics[width=0.495\textwidth]{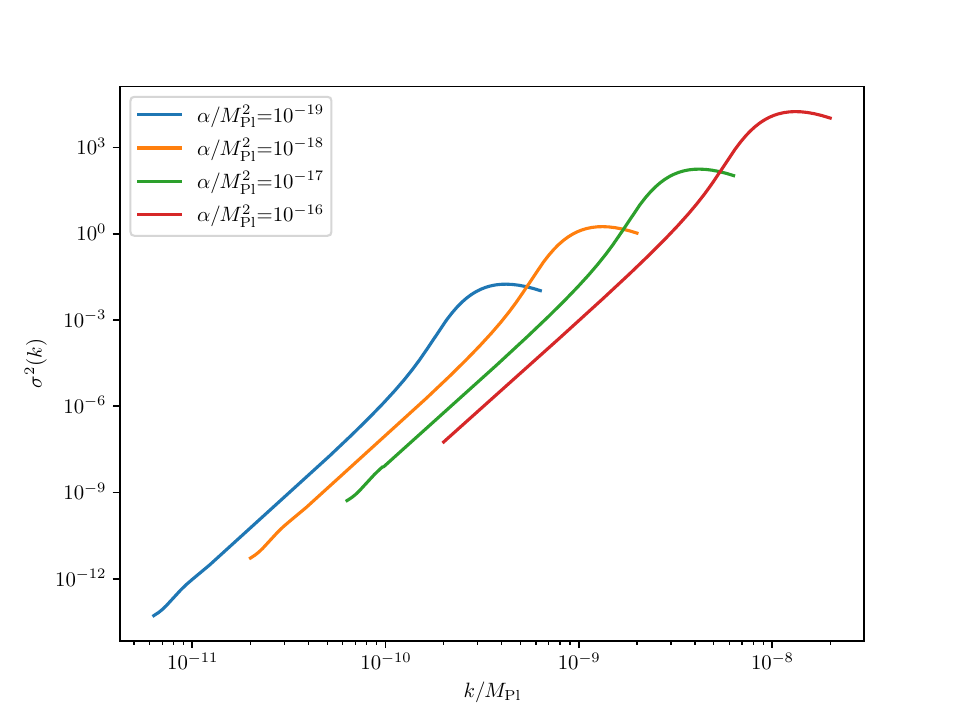}
\includegraphics[width=0.495\textwidth]{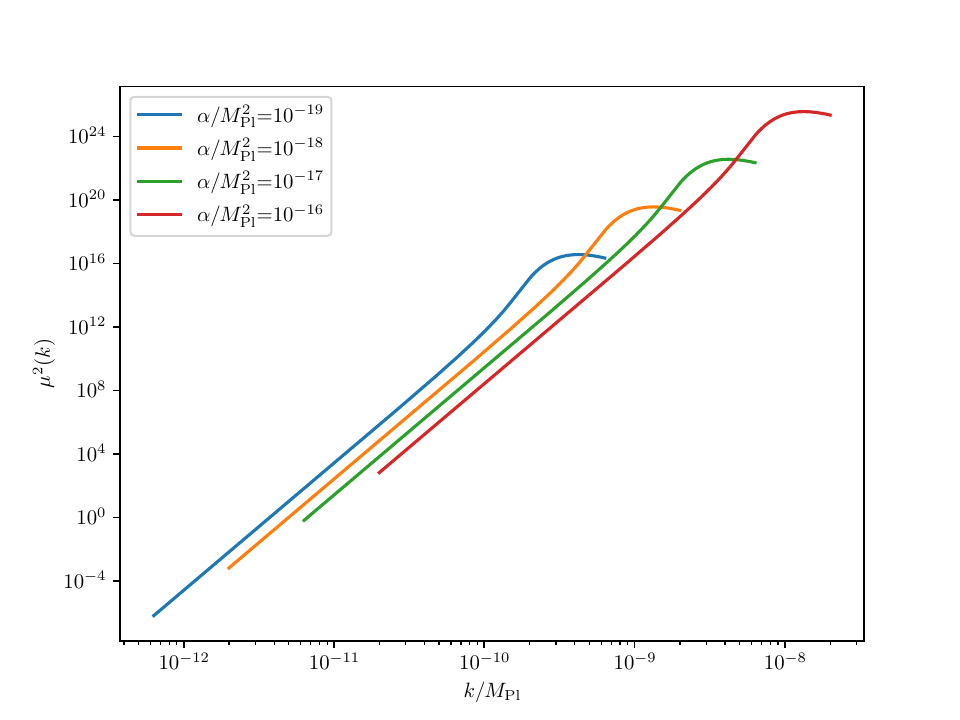}
\caption{{\it{The smoothed power spectra $\sigma^2(k)$ and $\mu^2(k)$ as a function of $k$ for different values of $\alpha$.}}}
\label{fig:sigma2_mu2_vs_alpha}
\end{center}
\end{figure*}

Writing now the fraction of the Universe at a peak of height $\nu\equiv\delta/\sigma$, which will collapse to form a PBH [See \Eq{eq:peak_density}] as a function of the energy density contrast one can recast it as

\beq
\beta_\delta = \frac{\mathcal{K}}{4\pi^2}\left(\delta-\frac{3\delta^2}{8}-\delta_\mathrm{c}\right)^\gamma\delta^3 e^{-
\frac{\delta^2}{2\sigma^2}}\frac{\mu^3}{\sigma^6}.
\eeq

\begin{figure}[h!]
\begin{center}
\includegraphics[width=0.495\textwidth]{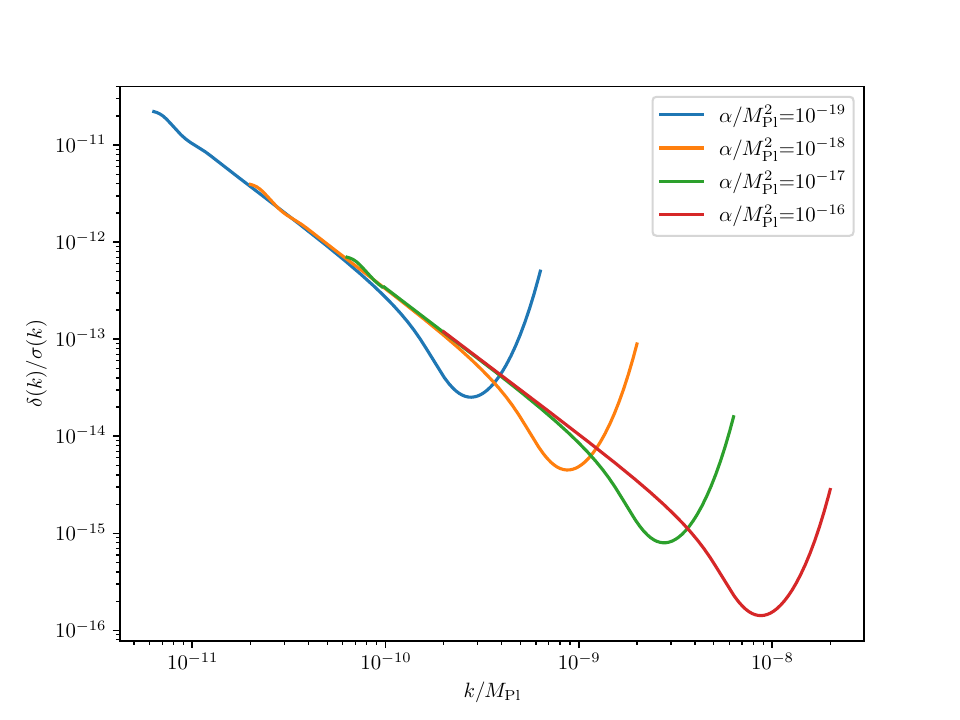}
\caption{{\it{$\delta(k)/\sigma(k)$ as a function of $k$ for different values of $\alpha$.}}}
\label{fig:delta_o_sigma}
\end{center}
\end{figure}

At the end, after integrating the $\beta_\delta$ over $\delta$ one will have that $\beta(k) = H(\sigma)\mu^3(k)/\sigma^6(k)$
with the function $H(\sigma)$ being defined as
\beq\label{eq:H_sigma}
H(\sigma)\equiv \int_{\delta_\mathrm{c,l-}}^{4/3}\frac{\mathcal{K}}{4\pi^2}\left(\delta-\frac{3\delta^2}{8}-\delta_\mathrm{c}\right)^\gamma\delta^3e^{-
\frac{\delta^2}{2\sigma^2}}
\eeq
As it was checked numerically [See \Fig{fig:delta_o_sigma}] for the range of $k$ values considered here $\delta/\sigma \ll 1$ and thus one can approximate $e^{-\frac{\delta^2}{2\sigma^2}} \simeq 1 - \frac{\delta^2}{\sigma^2}$. As we decrease $\sigma$, $H(\sigma)$ decreases as well. However, due to the $1/\sigma^6$ dependence of $\beta$, as we approach the region close to $k_\mathrm{max}$, we see the slight increase in $\beta(k)$ as can be seen in \Fig{fig:beta_vs_alpha}. This region where one observes this slight increase of the $\beta$ function can be roughly defined as $k>k_\mathrm{max}/4=\sqrt{\alpha}/2$.


\begin{figure}[h!]
\begin{center}
\includegraphics[width=0.495\textwidth]{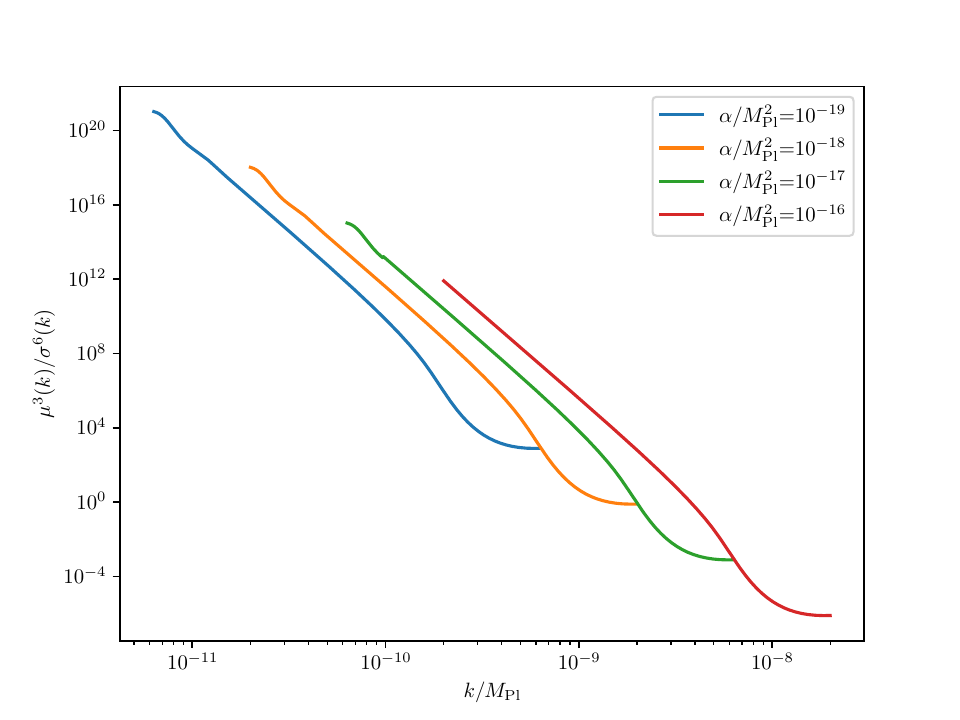}
\caption{{\it{The function $\mu^3(k)/\sigma^6(k)$ as a function of $k$ for different values of $\alpha$.}}}
\label{fig:mu3_sigma6_vs_alpha}
\end{center}
\end{figure}

\end{appendix}

\bibliographystyle{JHEP} 
\bibliography{PBH}
\end{document}